\documentclass[aps,10pt,tightenlines,notitlepage,showpacs,nofootinbib,longbibliography,eprint,superscriptaddress,twocolumn,eqsecnum,raggedbottom,preprintnumbers]{revtex4-2}

\usepackage[export]{adjustbox}
\usepackage{amsfonts}
\usepackage{amsmath}
\usepackage{amssymb}
\usepackage{array}
\usepackage{bbold}
\usepackage{bm}
\usepackage{booktabs}
\usepackage[dvipsnames,usenames]{color}
\usepackage{enumitem}
\usepackage{float}
\usepackage{graphicx}
\usepackage[utf8]{inputenc}
\usepackage{lipsum}
\usepackage{multirow}
\usepackage{nicefrac}
\usepackage[normalem]{ulem}
\usepackage[dvipsnames]{xcolor}
\usepackage{kantlipsum}
\usepackage{hyperref}
\hypersetup{
    colorlinks=true,
    urlcolor=blue,
    linkcolor = blue,
    urlcolor  = blue,
    citecolor = blue,
    anchorcolor = blue,
    pdftitle={JBW-CC-electroproduction},
    pdfauthor={JBW},
    pdfsubject={JBW-CC-electroproduction}}

\newcommand{\overbar}[1]{\mkern 1.5mu\overline{\mkern-1.5mu#1\mkern-1.5mu}\mkern 1.5mu}
\newcommand{\GeV}{{\rm GeV}}

\begin{document}


\title{Inclusion of \texorpdfstring{$K\Lambda$}~ electroproduction data in a coupled channel analysis}

\author{M.~Mai}
\email{mai@hiskp.uni-bonn.de}
\affiliation{Helmholtz-Institut f\"ur Strahlen- und Kernphysik (Theorie) and Bethe Center for Theoretical Physics,  Universit\"at Bonn, 53115 Bonn, Germany}
\affiliation{Institute for Nuclear Studies and Department of Physics, The George Washington University, Washington, DC 20052, USA}

\author{J.~Hergenrather}
\affiliation{Institute for Nuclear Studies and Department of Physics, The George Washington University, Washington, DC 20052, USA}

\author{M.~D\"oring}
\affiliation{Institute for Nuclear Studies and Department of Physics, The George Washington University, Washington, DC 20052, USA}
\affiliation{Thomas Jefferson National Accelerator Facility, Newport News, VA 23606, USA}

\author{T.~ Mart}
\affiliation{Departemen Fisika, FMIPA, Universitas Indonesia, Depok 16424, Indonesia}

\author{Ulf-G.~Mei{\ss}ner}
\affiliation{Helmholtz-Institut f\"ur Strahlen- und Kernphysik (Theorie) and Bethe Center for Theoretical Physics,  Universit\"at Bonn, 53115 Bonn, Germany}
\affiliation{Institute for Advanced Simulation and J\"ulich Center for Hadron Physics, Forschungszentrum J\"ulich,  52425 J\"ulich, Germany}
\affiliation{Tbilisi State University, 0186 Tbilisi, Georgia}

\author{D.~R\"onchen}
\affiliation{Institute for Advanced Simulation and J\"ulich Center for Hadron Physics, Forschungszentrum J\"ulich, 
52425 J\"ulich, Germany}

\author{R.~Workman}
\affiliation{Institute for Nuclear Studies and Department of Physics, The George Washington University, Washington, DC 20052, USA}

\collaboration{J\"ulich-Bonn-Washington Collaboration}
\preprint{JLAB-THY-23-3888}

\begin{abstract}
Exclusive electroproduction reactions provide an access to the  structure of excited baryons. To extract electroproduction multipoles encoding this information, the J\"ulich-Bonn-Washington (JBW) analysis framework is extended to the analysis of differential cross sections in $K\Lambda$ electroproduction. This update enlarges the scope of previous coupled-channel analyses of pions and eta mesons, with photoproduction reactions as boundary condition in all analyzed electroproduction reactions. Polarization observables are predicted and compared to recent CLAS data. The comparison shows the relevance of these data to pin down baryon properties.
\end{abstract}

\maketitle   

\section{Introduction}
\label{sec:introduction}
Electromagnetic probes of strongly interacting matter provide independent access to emergent phenomena of Quantum Chromodynamics (QCD) like resonances. 
Photoproduction reactions have been used to determine the spectrum and properties of excited baryons~\cite{Ireland:2019uwn, Thiel:2022xtb} as analyzed by different groups~\cite{Mart:2002gn, Shklyar:2005xg, Drechsel:2007if, Anisovich:2011fc, Workman:2012jf, Kamano:2013iva, Ronchen:2014cna, Hunt:2018mrt}. These analyses allow for a comparison to theory like lattice QCD~\cite{Burch:2006cc, Bulava:2010yg, Engel:2010my, Edwards:2011jj, Menadue:2011pd, Edwards:2012fx, Dudek:2012ag, Alexandrou:2013ata, Alexandrou:2015hxa, Stokes:2019zdd, RQCD:2022xux} or quark models~\cite{Ferraris:1995ui, Glozman:1995fu, 
Loring:2001kx, Giannini:2001kb, Santopinto:2004hw, Bijker:2009up, 
Liu:2022ndb}. See Ref.~\cite{Mai:2022eur} for a recent review. Notably, first calculations of meson-baryon scattering amplitudes in lattice QCD have appeared recently, some of them containing the $\Delta(1232)3/2^+$ resonance~\cite{Meissner:2010ij,Andersen:2017una, Silvi:2021uya, Pittler:2021bqw, Bulava:2022vpq}.
Complementary to photoproduction reactions, radiative decays of excited baryons, such as measured by CLAS~\cite{CLAS:2005bgo}, can reveal information about their nature, see, e.g., Refs.~\cite{Geng:2007hz, Doring:2007rz, Doring:2006ub}. 

In addition, the momentum transfer of the probe can be tuned once the photon is allowed to become virtual, testing strong interactions at different scales. Indeed, electroproduction reactions are a prime tool to study the structure of excited baryons~\cite{Mokeev:2022xfo, Ramalho:2023hqd}. One cannot directly test the response of a resonance to a virtual photon, but  determine transition form factors in the electro-excitation of the resonance from the nucleon. One can map out the transverse charge density by using electromagnetic form factors~\cite{Carlson:2007xd}. The $Q^2$-dependent multipoles can also be used to test chiral perturbative calculations~\cite{Bernard:1992ms, Bernard:1992rf, Bernard:1993bq, Bernard:1994dt, Bernard:1996bi, Steininger:1996xw, Bernard:2000qz, Krebs:2004ir} and  unitary extensions~\cite{Jido:2007sm, Doring:2010rd, Mai:2020ltx}, chiral resonance calculations~\cite{Gail:2005gz, Bauer:2014cqa}, and quark models~\cite{
Merten:2002nz, Gross:2006fg, 
Ramalho:2011ae, Santopinto:2012nq, Golli:2013uha, Aznauryan:2017nkz, Obukhovsky:2019xrs, Ramalho:2020nwk}.
Notably, a gauge invariant chiral unitary framework for kaon electroproduction was developed in Ref.~\cite{Borasoy:2007ku} and extended later~\cite{Ruic:2011wf, Mai:2012wy}. Transition form factors also serve as point of comparison for dynamical quark calculations referred to as Dyson-Schwinger approaches~\cite{Cloet:2008re, Wilson:2011aa, 
Segovia:2015hra, Eichmann:2016hgl, Chen:2018nsg, Qin:2019hgk}.
In this context, remarkable agreement of the lower-lying baryon spectrum with predictions has been achieved~\cite{Eichmann:2016hgl, Qin:2019hgk}, showing little evidence for a ``missing resonance'' problem at lower energies. See Refs.~\cite{Aznauryan:2011qj, Aznauryan:2012ba, Bashir:2012fs, Eichmann:2016yit, Eichmann:2022zxn} for reviews.
Methods to study the $Q^2$-dependence of resonance couplings in lattice QCD were proposed in Ref.~\cite{Agadjanov:2014kha}.
A pioneering lattice calculation was carried out recently in the meson sector~\cite{Radhakrishnan:2022ubg}. 

Transition form factors have been defined in different ways~\cite{Aznauryan:2011qj}, but the only reaction-independent definition is given in terms of $Q^2$-dependent couplings at the resonance pole, to be determined by an analytic continuation of electroproduction multipoles~\cite{Tiator:2016btt}. 

The multipoles themselves are determined by analyzing the exclusive electroproduction of one or more mesons. The advantage of simultaneously analyzing different final states in a coupled-channel approach lies in the factorization of the amplitude at the pole, i.e., the fact that the resonance transition form factor is the same for any final state. 

Another reason to perform global analyses of electroproduction reactions is the need to analyze as many data simultaneously as possible. The data situation in electroproduction reactions tends to be more challenging than in photoproduction. On one hand, this is due to the presence of another kinematic variable in addition to the energy $W$, namely the virtuality of the photon $Q^2=-q^2$, where $q$ is the transferred four-momentum of the photon. Even though the number of data points is larger in electro- than in photoproduction, the data are still sparser due to this additional variable. On the other hand, there are longitudinal multipoles to be determined from data, in addition to the electric and magnetic ones that parameterize the photoproduction amplitudes. The related question of how many measurements are necessary to determine a truncated partial-wave expansion of the electroproduction amplitude is discussed in Refs.~\cite{Tiator:2017cde, Wunderlich:2021xhp}.

All this motivates the inclusion of $K\Lambda$ electroproduction reported in this paper. This coupled-channel extension is based on previous analyses within the J\"ulich-Bonn-Washington (JBW) framework of pion~\cite{Mai:2021vsw} and eta-meson~\cite{Mai:2021aui} electroproduction. Representing the first coupled-channel electroproduction analysis, data at the photon point ($Q^2=0$) are also included as a boundary condition from previous analysis of pion~\cite{Ronchen:2014cna}, eta~\cite{Ronchen:2015vfa}, and $K\Lambda$~\cite{Ronchen:2018ury} photoproduction. The model was recently extended to $K\Sigma$ photoproduction~\cite{Ronchen:2022hqk} and pion-induced $\omega$ productions~\cite{Wang:2022osj}, but the  analysis presented here is based on the JüBo2017 solution that includes $\pi N,\,\eta N$, and $K\Lambda$ photoproduction. In addition, the coupled-channel amplitude was also used to simultaneously analyze the pion-induced production of the aforementioned meson-baryon states~\cite{Doring:2010ap, Ronchen:2012eg}, providing additional constraints on the strong final-state interactions in both photo- and electroproduction. The comparisons of data and fit solutions of pion- and real-photon-induced reactions (J\"uBo) have been collected on a website~\cite{JB-homepage}. The JBW electroproduction solutions are collected on another interactive website~\cite{JBW-homepage}.

The single-channel analysis of single-meson electroproduction data has a long history; one of the first approaches is MAID for pion photo- and electroproduction~\cite{Tiator:2003uu, Drechsel:2007if}, later complemented by a chiral-MAID approach at low energies~\cite{Hilt:2013fda}. There is also the etaMAID2001 analysis on eta electroproduction~\cite{Chiang:2001as}. See Ref.~\cite{Tiator:2011pw} for a review. The CLAS collaboration extracted helicity amplitudes for several resonances from their experiment~\cite{CLAS:2009ces}, including the unusual zero in the $A_{\nicefrac{1}{2}}$ Roper form factor~\cite{Burkert:2017djo, Burkert:2022ioj}; see also 
Refs.~\cite{Aznauryan:2002gd,CLAS:2008roe} for other CLAS analyses. The ANL-Osaka group analyzed electroproduction data in the context of neutrino-induced reactions~\cite{Nakamura:2015rta}. Questions on efficient parametrizations of electroproduction amplitudes and transition form factors are discussed in Refs.~\cite{Ramalho:2017muv, Ramalho:2017xkr, Ramalho:2019ocp}.
The two-pion electroproduction reaction has also been measured at CLAS and analyzed with the JM reaction model~\cite{CLAS:2012wxw, Mokeev:2015lda, Mokeev:2020hhu, Mokeev:2023zhq}, see also Ref.~\cite{Bernard:1994ds}. Notably, much higher $Q^2$ values for resonance transition form factors become accessible in ongoing CLAS12 experiments~\cite{Aznauryan:2012ba}.

Most relevant for the present analysis of $K\Lambda$ electroproduction is KAON-MAID~\cite{Bennhold:1999mt, Mart:2002gn}, an analysis using an effective Lagrangian approach~\cite{Maxwell:2012zz}, and the more recent analyses using a Regge-plus-resonance (RPP) amplitude~\cite{Corthals:2007kc, DeCruz:2012bv}. See Ref.~\cite{Carman:2018fsn} for an overview of kaon electroproduction reactions and Refs.~\cite{Haberzettl:1998aqi, Mart:1999ed, Mart:2006dk, Mart:2011ez, JPAC:2016lnm, Blin:2021twt, Haberzettl:2021wcz} for related analyses and theoretical developments by JPAC and others.

In the present analysis we fit mostly cross section data for $\gamma^* p\to K\Lambda$ from CLAS. Notably, the data base was recently enlarged through the addition of beam-recoil transfer polarization data~\cite{CLAS:2022yzd}. In the presented update of the JBW approach, we predict the latter data but do not fit them. This allows for a check of how much they will constrain the multipole extraction in future analyses, for which $K\Sigma$ electroproduction data should also be included. As $K\Lambda$ electroproduction requires an extension to higher energies, we enlarge the range of analyzed electroproduction data accordingly from $W=1.6$~GeV to $W=1.8$~GeV, compared to the previous study~\cite{Mai:2021aui}. Also, the range in $Q^2$ was extended up to $Q^2=8\,\GeV^2$ for all analyzed final states, and we also include F-waves now, owing to the higher energy range. The extraction of resonance transition form factors is scheduled once the analysis stage is complete.

This study is organized as follows. Section~\ref{sec:formalism} outlines formal aspects of the JBW approach to pseudo-scalar meson electroproduction. Furthermore, we define the parametrization of the  $Q^2$ dependence connected to the photon point at $Q^2=0$, where the underlying J\"uBo model describes photon- and meson-induced reactions. 
Section~\ref{sec:expdata} describes the data and fit procedures with different data weighting. Section~\ref{sec:results} compares our fits to data and puts the analysis in context.

\section{Formalism}
\label{sec:formalism}

\begin{figure}[t]
\centering
\includegraphics[width=\linewidth,trim=0cm 6cm 12cm 6cm,clip]{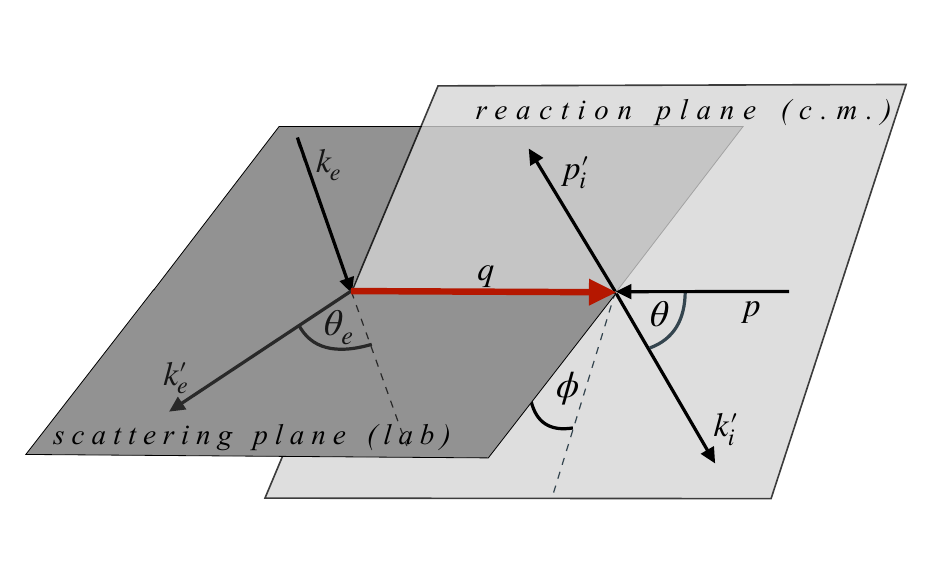}
\caption{
\label{fig:kinematics}
Kinematics of an electroproduction experiment with the final meson-baryon state $i$. The scattering plane is defined by the respective in/outgoing electron momenta $k_e/k_e'$ with the electron scattering angle $\theta_e$. The reaction plane is spanned by the virtual photon and the outgoing meson, scattered by an angle $\theta$. The momenta $q$ and $p$ correspond to the virtual photon and target
nucleon while $k_i'$ and $p_i'$ correspond to the outgoing meson and baryon, respectively.
}
\end{figure}

We summarize the formalism following closely Refs.~\cite{Mai:2021vsw,Mai:2021aui} in a more general form. The multichannel meson electroproduction process under consideration is
\begin{align}
\gamma^*(\bm{q})+p(\bm{p})
\to
M(\bm{k}'_i)+B(\bm{p}'_i)\,,
\end{align}
where bold symbols denote three-momenta  throughout the manuscript. The meson and baryon in the final state, with the index $i$, are denoted by $M$ and $B$, respectively. As shown in Fig.~\ref{fig:kinematics}, the process occurs in two steps, with a virtual photon $\gamma^*(\bm{q})$ being produced via $e_{\rm in}(\bm{k}_e)\to e_{\rm out}(\bm{k}'_e)+\gamma^*(\bm{q})$, which then scatters off the proton to a final meson-baryon state. The momentum transfer $Q^2=-\omega^2+\bm{q}^2$, where $\omega$ is the photon energy, is non-negative for spacelike processes, and acts as an independent kinematical variable in addition to the total energy in the center-of-mass (c.m.) frame, $W$. In this frame, the magnitude of the three-momentum of the photon ($q=|\bm{q}|$) and produced meson ($k'_i=|\bm{k}_i'|$) read
\begin{align}
q=\frac{\sqrt{\lambda(W^2,m_p^2,-Q^2)}}{2W}\,,~
k_i'=\frac{\sqrt{\lambda(W^2,m_i^2,M_i^2)}}{2W}\,,
\label{eq:qk}
\end{align}
where $\lambda(x,y,z)=x^2+y^2+z^2-2xy-2yz-2zx$ denotes the usual K\"all\'en triangle function. Meson and baryon masses are denoted  by $M$ and $m$, respectively.
With two incoming and three outgoing particles there are $(3+2)\times3-10=5$ independent kinematic variables. The canonical choice for the remaining three (in addition to $W$ and $Q^2$) variables is illustrated in Fig.~\ref{fig:kinematics}. The quantity $\epsilon$ is defined through
\begin{align}
\label{eq:epsilon}
   \epsilon=\frac{1}{1+2({q_L^2}/{Q^2})\tan^2{\theta_e}/{2}}\,,
\end{align}
and contains the electron scattering angle $\theta_e$ and $q_L$ denotes the photon three-momentum in the laboratory frame. The angle of the reaction plane to the scattering plane is given by $\phi$, and $\theta$ is the c.m.\ meson scattering angle in the latter plane. The experimental data discussed in Sec.~\ref{sec:expdata}, symbolized as $O$, are represented with respect to these five variables, i.e., $O(Q^2,W,\phi,\theta,\epsilon)$.

As discussed in the previous paper~\cite{Mai:2021vsw}, based on the seminal works~\cite{Chew:1957tf, Dennery:1961zz,Berends:1967vi,CiofiDegliAtti:1981wk}, the process of a  photon-induced production of a meson off a nucleon is encoded in the transition amplitude. In the one-photon approximation, and considering the continuity equation for the current, the latter can be expressed in terms of three independent multipoles for a fixed quantum number $\ell_\pm$ of the final meson-baryon state. We chose those to be electric, magnetic and longitudinal multipoles $E_{\ell\pm}^\mu(W,Q^2)$, $M_{\ell\pm}^\mu(W,Q^2)$ and $L_{\ell\pm}^\mu(W,Q^2)$ with the latter related to the often-used Coulomb multipole as $\omega C_{\ell\pm}(W,Q^2)=q L_{\ell\pm}(W,Q^2)$. Each of these multipoles carries a discrete index corresponding to the total angular momentum $J=\ell\pm 1/2$ and final-state index $\mu$, e.g., $E^{\eta p}_{0+}$.

We construct the electroproduction multipoles on the basis of the dynamical coupled-channel J\"ulich-Bonn (J\"uBo) approach~\cite{Ronchen:2012eg,Ronchen:2014cna} that provides the boundary condition at $Q^2=0$, incorporating the experimental information from real-photon and pion-induced reactions. In this approach, two-body unitarity and analyticity are respected and the baryon resonance spectrum is determined in terms of poles in the complex energy plane on the second Riemann sheet~\cite{Doring:2009yv, Doring:2009bi}. In particular, we use the J\"ubo2017 solution that includes $\pi N$, $\eta N$, and $K\Lambda$ photoproduction~\cite{Ronchen:2018ury}.

Extending the ansatz of the J\"uBo approach, we begin by introducing a generic function ($\bar{\cal{M}}$) for each electromagnetic multipole  (${\cal M}_{\mu\gamma^*}\in \{E^\mu,M^\mu,L^\mu\}$) as
\begin{align}
\label{eq:m_electro}
    \overbar{\cal M}_{\mu\gamma^*}&(k,W,Q^2)=V_{\mu\gamma^*}(k,W,Q^2)\\
    +&\sum_\kappa\int\limits_0^\infty dp\, p^2\, T_{\mu\kappa}(k,p,W)G_\kappa(p,W)V_{\kappa\gamma^*}(p,W,Q^2)\,,\nonumber
\end{align}
where $\mu$ is a channel index and the summation extends over intermediate meson-baryon channels $\kappa\in\{\pi N,\eta N, K\Lambda, K\Sigma, \pi\Delta, \rho N\}$. Note that the $\sigma N$ channel is not part of this list. The $\sigma N$ channel is part of the final-state interaction, but neither the hadronic resonance vertex functions nor the photon is directly coupled to it. However, once photo- or electroproduction data of the $\pi\pi N$ final state are analyzed, such couplings will become relevant and will be included.
Note that we have suppressed isospin and the angular momentum index $\ell_\pm$ in Eq.~\eqref{eq:m_electro}.

The electroproduction kernel $V_{\mu\gamma^*}$ in Eq.~\eqref{eq:m_electro} is parameterized as
\begin{align}
\label{eq:v_electro}
    V_{\mu\gamma^*}(p,W,Q^2)=&
    \alpha^{NP}_{\mu\gamma^*}(p,W,Q^2)\\\nonumber
    &+\sum_{i=1}^{i_\text{max}}\frac{\gamma^a_{\mu;i}(p)\gamma^c_{\gamma^*;i}(W,Q^2)}{ W-m^b_i}\,,
\end{align}
introducing the $Q^2$-dependence via a separable ansatz,
\begin{align}
    \alpha^{NP}_{\mu\gamma^*}(p,W,Q^2) &={\tilde F}_\mu(Q^2)\alpha^{NP}_{\mu\gamma}(p,W)\,,\nonumber\\
    \gamma^c_{\gamma^*;i}(W,Q^2)       &={\tilde F}_i(Q^2)\gamma^c_{\gamma;i}(W)\,.
\label{eq:ff_electro_2}
\end{align}
The $Q^2$-independent pieces on the right-hand side of both equations represent the input from the J\"uBo2017 solution~\cite{Ronchen:2018ury}. Specifically, $\gamma^c_{\gamma;i}$ describes the interaction of the photon with the resonance state $i$ with bare mass $m_i^b$ and $\alpha^{NP}_{\mu\gamma}$ accounts for the coupling of the photon to the so-called background or non-pole part of the amplitude. Both quantities are parameterized by energy-dependent polynomials, see Ref.~\cite{Ronchen:2014cna}.

The $Q^2$-dependence is encoded entirely in the channel-dependent form-factor ${\tilde F}_\mu(Q^2)$ and another channel-independent form-factor ${\tilde F}_i(Q^2)$ that depends on the resonance index $i$. We emphasize that this structure is inherited from the J\"uBo photoproduction ansatz, which separates the photon-induced vertex ($\gamma^c$) from the decay vertex of an s-channel resonance to the  meson-baryon pair ($\gamma_\mu^a$). Both ${\tilde F}_\mu(Q^2)$ and ${\tilde F}_i(Q^2)$ are chosen as
\begin{align}
    {\tilde F}_\mu(Q^2)&={\tilde F}_D(Q^2)\,e^{-\beta_\mu^0 Q^2/ m^2}\,P^N(Q^2/m^2,\vec{\beta}_\mu)\,,\nonumber\\
    {\tilde F}_i(Q^2)&={\tilde F}_D(Q^2)\,e^{-\delta_i^0 Q^2/ m^2}\,P^N(Q^2/m^2,\vec{\delta}_i)\,,
    \label{eq:formfactor-Ftilde}
\end{align}
where $P^N(x,\vec y)= 1 + xy_1 +...+x^Ny_N$ is a general polynomial with free parameters to be fitted together with $\delta_i^0$ and $\beta^0_\mu$ to the  electroproduction data. The parameter-free form factor ${\tilde F}_D(Q^2)$ encodes the empirical dipole behavior, usually implemented in such problems, as well as a Woods-Saxon form factor which ensures suppression at large $Q^2$. It reads
\begin{align}
    \label{eq:formfactor-FD}
    {\tilde F}_D(Q^2)=\frac{1}{(1+Q^2/b^2)^{2}}\,
    \frac{1+e^{-Q_r^2/Q_w^2}}{1+e^{(Q^2-Q_r^2)/Q_w^2}}
\end{align}
with $b^2=0.71$~GeV${}^2$, $Q_w^2=0.5~{\rm GeV}^2$ and  $Q_r^2=10.0~{\rm GeV}^2$, see Ref.~\cite{Mai:2021vsw} for more details. Note that we have increased the range parameter $Q_r^2$, such that  the suppression from the Wood-Saxon form factor is only relevant beyond the range of data considered here ($Q^2>8\,\GeV^2$).

As stated above, this procedure relies heavily on the input from the photoproduction, i.e., the functions $\alpha^{NP}_{\mu \gamma}(p,W)$ and $\gamma^c_{\gamma;i}(W)$. This input does not exist for the longitudinal multipoles as their contribution vanishes exactly at the photon-point. In this case we employ a strategy similar to that of Ref.~\cite{Mai:2012wy}:

1) We recall that at the \emph{pseudo-threshold} ($q=0$) the electric and longitudinal multipoles are related according to Siegert's condition~\cite{Siegert:1937yt, Tiator:2016kbr} as
\begin{equation}
    \frac{E_{\ell+}}{L_{\ell+}}\Bigg|_{q=0}=1\,,\qquad
    \frac{E_{\ell-}}{L_{\ell-}}\Bigg|_{q=0}=\frac{\ell}{1-\ell}\,.
\label{eq:Siegerts_condition}
\end{equation}
For more details, see Sec.~2.2-2.3 of Ref.~\cite{Mai:2012wy}, or the earlier derivations in Refs.~\cite{CiofiDegliAtti:1981wk,Tiator:2016kbr}. Therefore, we apply at the nearest  pseudo-threshold point, $Q^2_{\rm PT}=-(W-m)^2$,
\begin{align}
\alpha^{NP,L_{\ell\pm}}_{\mu\gamma^*}(p,W,Q^2)&=
\frac{\omega}{\omega_{\rm PT}}
\frac{{\tilde F}_D(Q^2)}{{\tilde F}_D(Q^2_{\rm PT})}
\label{pt_cond3}\\
&\times D_\mu^{\ell\pm}(W,Q^2)
\alpha^{NP,E_{\ell\pm}}_{\mu\gamma^*}(p,W,Q^2_{\rm PT})\,,\nonumber
\end{align}
and
\begin{align}
\gamma^{c,L_{\ell\pm}}_{\gamma^*;i}(W,Q^2)=
\frac{\omega}{\omega_{\rm PT}}&
\frac{{\tilde F}_D(Q^2)}{{\tilde F}_D(Q^2_{\rm PT})}
\label{pt_cond4}\\
&\times\tilde D_i^{\ell\pm}(W,Q^2)
\gamma^{c,E_{\ell\pm}}_{\gamma^*;i}(W,Q^2_{\rm PT})\,.\nonumber
\end{align}
The photon energy is $\omega_{\rm PT}=(W^2-m^2-Q_{\rm PT}^2)/(2W)$. The new functions $D^{\ell\pm}(Q^2)$ ensure Siegert's condition and a consistent falloff behavior in $Q^2$ as
\begin{align}
D_\mu^{\ell+}(W,Q^2)&=
    e^{-\beta_\mu^0 q/q_\gamma}\,P^{N}(q/q_\gamma,\vec\beta_\mu)\,,\label{eq:D-formfactor}\\
\tilde D_i^{\ell+}(W,Q^2)&=
    e^{-\delta_i^0 q/q_\gamma}\,P^{N}(q/q_\gamma,\vec\delta_i)\,,\nonumber\\
D_\mu^{\ell-}(W,Q^2)&=
    -\frac{\ell-1}{\ell}e^{-\beta_i^0 q/q_\gamma}\,P^{N}(q/q_\gamma,\vec\beta_\mu)\,,\nonumber\\
\tilde D_i^{\ell-}(W,Q^2)&=
    -\frac{\ell-1}{\ell}e^{-\delta_i^0 q/q_\gamma}\,P^{N}(q/q_\gamma,\vec\delta_i)\,,\nonumber
\end{align}
respectively, to the pole and non-pole part for $q_\gamma=q(Q^2=0)$.

2) In two specific cases ($(\ell\pm,I)=(1-,1/2)$ and $(\ell\pm,I)=(1-,3/2)$) the electric multipole vanishes due to selection rules, rendering the implementation of Siegert's theorem nonsensical. In these cases, we decided to obtain the longitudinal multipole from the magnetic one using a new real-valued normalization constants $\zeta^{NP}$ to be determined from the fit,
\begin{align}
\label{eq:zetanp}
\alpha^{NP,L_{\ell\pm}}_{\mu\gamma^*}(p,W,Q^2)&=\zeta^{NP}_\mu\frac{\omega}{\omega_{\rm PT}}{\tilde F}^\mu(Q^2)\\
&\qquad\times \alpha^{NP,M_{\ell\pm}}_{\mu\gamma^*}(p,W)\,,\nonumber\\
\gamma^{c,L_{\ell\pm}}_{\gamma^*;i}(W,Q^2)&=\zeta_i\frac{\omega}{\omega_{\rm PT}}{\tilde F}^\mu(Q^2)\gamma^{c,M_{\ell\pm}}_{\gamma;i}(W)\,.\nonumber
\end{align}
Using the magnetic multipole as starting point, and a real-valued normalization constant ensures that Watson's theorem is fulfilled. 
\begin{figure*}
    \centering
    \includegraphics[width=0.42\linewidth]{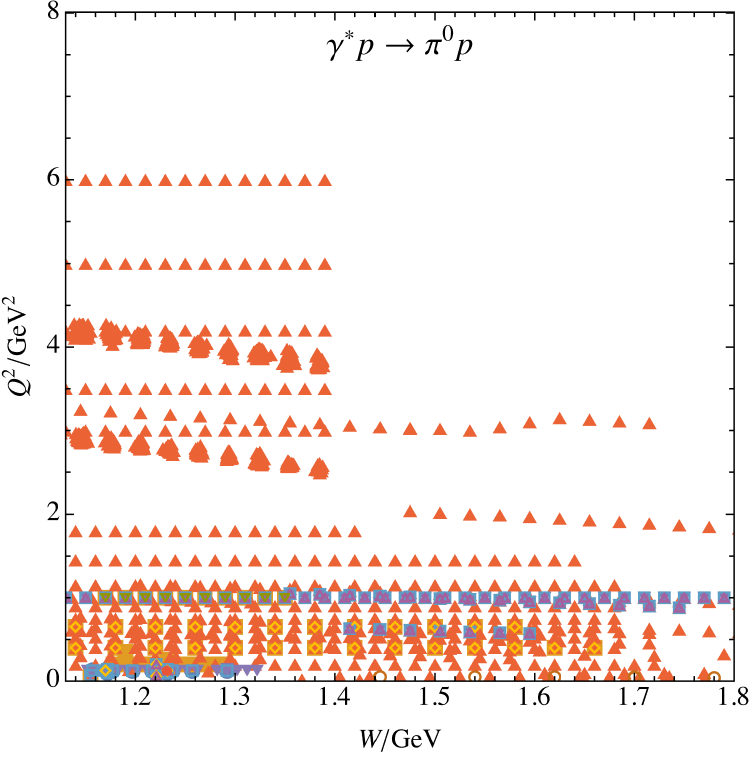}
    ~~~
    \includegraphics[width=0.42\linewidth]{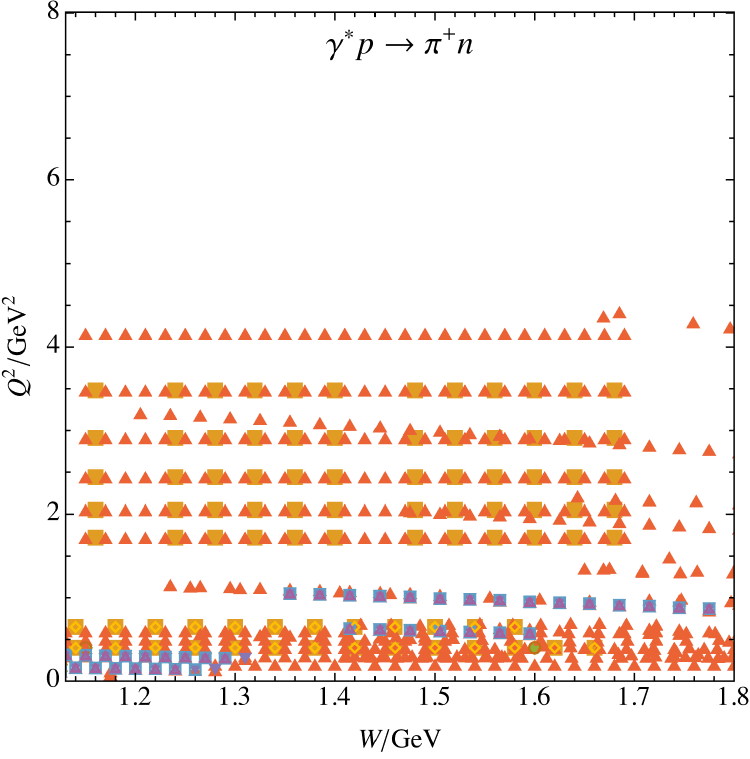}\\
    \includegraphics[width=0.42\linewidth]{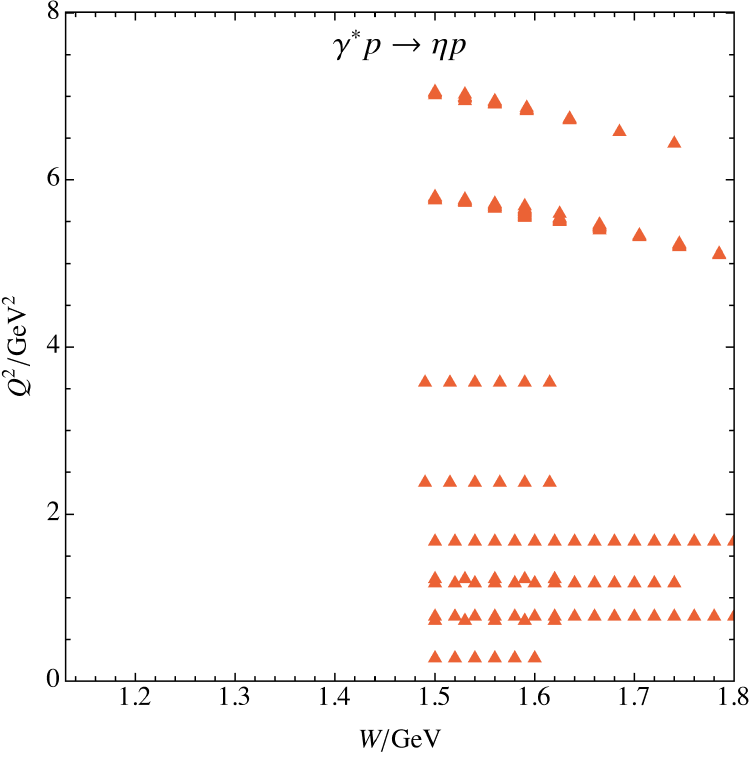}
    ~~~
    \includegraphics[width=0.42\linewidth]{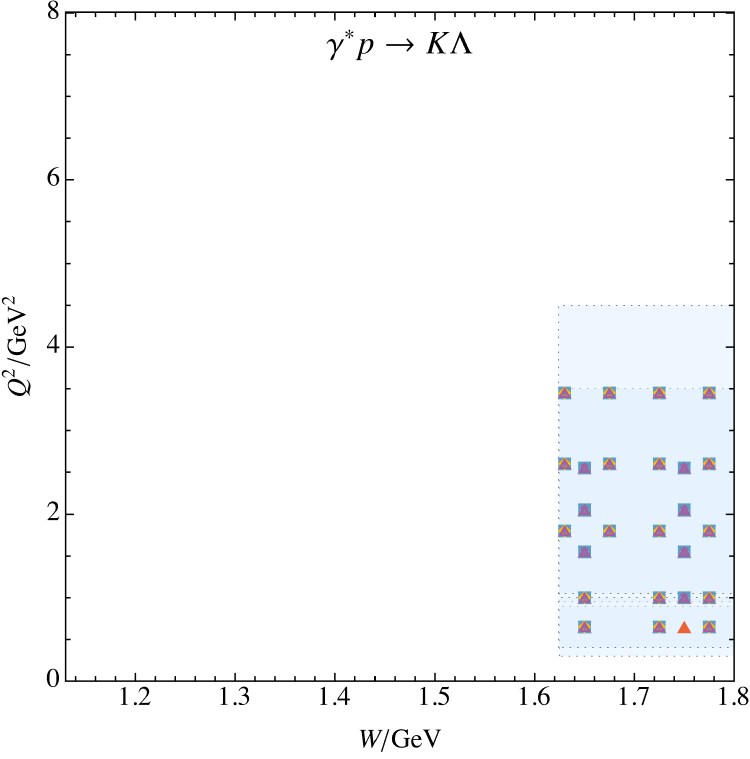}\\~\\
\begin{tabular}{|
c p{.07\linewidth}|
p{.30\linewidth}|p{.3\linewidth}|p{.13\linewidth}|p{.15\linewidth}|}
    \hline
    \multicolumn{2}{|c|}{Type}&$N_{\rm data}^{\pi^0p}$&$N_{\rm data}^{\pi^+n}$&$N_{\rm data}^{\eta p}$&$N_{\rm data}^{K\Lambda}$\\
    \hline
    \hline
    ${\color{Cerulean}\bullet}$&$\rho_{LT}$
    &45~\cite{Mertz:1999hp,Elsner:2005cz}
    &--
    &--
    &--\\
    ${\color{Dandelion}\blacksquare}$&$\rho_{LT'}$
    &2768~\cite{Joo:2003uc, Sparveris:2002fh, Kelly:2005jj, Bartsch:2001ea,Bensafa:2006wr}
    &5068~\cite{Joo:2004mi, PARK-pc-08-2007}
    &-- 
    &-- \\
    ${\color{ForestGreen}\blacklozenge}$&$\sigma_L$
    &--
    &2~\cite{Gaskell:2001fn}
    &--
    &--\\
    ${\color{BurntOrange}\blacktriangle}$&$d\sigma/d\Omega$
    &48135~\cite{Laveissiere:2003jf, Ungaro:2006df, Gayler:1971zz, May:1971zza, Frolov:1998pw, Hill:1977sy, Joo:2001tw, Siddle:1971ug, Haidan:1979yqa, Sparveris:2002fh, Kelly:2005jj, Kalleicher:1997qf, Baetzner:1974xy, Latham:1979wea, Latham:1980my, Stave:2006jha, Sparveris:2006uk, Alder:1975xt, Afanasev:1975qa, Shuttleworth:1972nw, Blume:1982uh, Rosenberg:1979zm, Gerhardt:1979zz, MONTANA-phd1971}
    &44266~\cite{Egiyan:2006ks, Breuker:1977vy, PARK-pc-08-2007, Bardin:1975oea, Bardin:1977zu, Gerhardt:1979zz, DavenportMartyn1980Eona, Vapenikova:1988fd, Hill:1977sy, Alder:1975na, Evangelides:1973gg, Breuker:1982nw, Breuker:1982um, Bebek:1974ww, Brown:1973wr, Litt:1971xx}
    &3665~\cite{JeffersonLabE94014:1998czy, CLAS:2007bvs, CLAS:2000mbw, Dalton:2008aa}
    &2055~\cite{Carman:2012qj, CLAS:2006ogr}\\
    ${\color{CadetBlue}\blacktriangledown}$&$\sigma_T+\epsilon \sigma_L$
    &384~\cite{Laveissiere:2003jf, Mertz:1999hp, Sparveris:2002fh, Kunz:2003we, Stave:2006ea, Sparveris:2006uk, Sparveris:2004jn, Alder:1975xt, Laveissiere:2003jf}
    &182~\cite{Breuker:1977vy, Alder:1975na}
    &--
    &204~\cite{Carman:2012qj, CLAS:2006ogr}\\
    ${\color{Maroon}\circ}$&$\sigma_{T}$
    &30~\cite{Blume:1982uh}
    &2~\cite{Gaskell:2001fn}
    &--
    &--\\
    ${\color{Cerulean}\square}$&$\sigma_{LT}$
    &373~\cite{Laveissiere:2003jf, Sparveris:2002fh, Mertz:1999hp, Kunz:2003we, Stave:2006ea, Sparveris:2006uk, Sparveris:2004jn, Alder:1975xt, Laveissiere:2003jf}
    &138~\cite{Breuker:1977vy, Alder:1975na}
    &--
    &204~\cite{Carman:2012qj, CLAS:2006ogr}\\
    ${\color{Dandelion}\Diamond}$&$\sigma_{LT'}$
    &214~\cite{Joo:2003uc, Kunz:2003we, Stave:2006ea, Sparveris:2006uk}
    &208~\cite{Joo:2003uc}
    &--
    &156~\cite{Carman:2012qj, CLAS:2008agj}\\
    ${\color{Mulberry}\triangle}$&$\sigma_{TT}$
    &327~\cite{Laveissiere:2003jf, Stave:2006ea, Sparveris:2006uk, Sparveris:2004jn, Alder:1975xt, Laveissiere:2003jf}
    &123~\cite{Breuker:1977vy, Alder:1975na}
    &--
    &204~\cite{Carman:2012qj, CLAS:2006ogr}\\
    ${\color{JungleGreen}\nabla}$&$K_{D1}$
    &1527~\cite{Kelly:2005jj}
    &--
    &--
    &--\\
    ${\color{BrickRed}\bullet}$&$P_Y$
    &--
    &2~\cite{Warren:1999pq, Pospischil:2000ad}
    &--
    &--\\
    \hline\hline
    \multicolumn{2}{|c|}{Total}
    &53804
    &49989
    &3665
    &2823\\
    \hline
\end{tabular}
    \caption{Overview over the fit ranges and data types used in this work. The kinematical region covered by the recent beam-recoil transferred polarization measurement of Ref.~\cite{CLAS:2022yzd} is represented by the blue shaded area. These data are not part of the fits but is discussed in Sec.~\ref{sec:results}.}
    \label{fig:DATA}   
\end{figure*}
Before writing down the final relation between the generic multipole functions ($\bar E_{\ell\pm}$, $\bar M_{\ell\pm}$, $\bar L_{\ell\pm}$) and corresponding multipoles, we note that the latter obey a certain behavior at the pseudo- ($q=0$) and production threshold ($k=0$),
\begin{align}
&\ell\geq0: \quad\lim_{k\to 0}E_{\ell+}=k^\ell~, &&\lim_{q\to 0}E_{\ell+}=q^\ell~,\nonumber\\
&\ell\geq0: \quad\lim_{k\to 0}L_{\ell+}=k^\ell~, &&\lim_{q\to 0}L_{\ell+}=q^\ell~,\nonumber\\
&\phantom{\ell=1:}    \quad\lim_{k\to 0}L_{1-}=k~, &&\lim_{q\to 0}L_{1-}=q~,\nonumber\\
&\ell\geq1: \quad\lim_{k\to 0}M_{\ell\pm}=k^\ell~, &&\lim_{q\to 0}M_{\ell\pm}=q^\ell~,\nonumber\\
&\ell\geq2: \quad\lim_{k\to 0}E_{\ell-}=k^\ell~, &&\lim_{q\to 0}E_{\ell-}=q^{\ell-2}~,\nonumber\\
&\ell\geq2: \quad\lim_{k\to 0}L_{\ell-}=k^\ell~, &&\lim_{q\to 0}L_{\ell-}=q^{\ell-2}~.
\label{eq:pseudo-thre-conditions}
\end{align}
We incorporate these conditions using 
\begin{align}
    {\cal M}_{\mu\gamma^*}(k,W,Q^2)=
    R_{\ell'}(\lambda, q/q_\gamma)\overbar{\cal M}_{\mu\gamma^*}(k,W,Q^2)
    \label{ampl_2}
\end{align}
for each multipole type and total angular momentum individually. Here,
\begin{align}
    R_{\ell'}(\lambda,r)&=\frac{B_{\ell'}(\lambda r)}{B_{\ell'}(\lambda)}
    \\
    &\text{with}\quad\ell'=
\left\{
\begin{matrix}
\ell,~~\text{for}~E_{\ell+},L_{\ell\pm},M_{\ell\pm}~,~~~\\
\ell-2,~~\text{for}~E_{\ell-},L_{\ell-}\text{  and  }\ell\geq 2\,,
\end{matrix}
\right. \nonumber
\end{align}
in terms of the  Blatt-Weisskopf barrier-penetration factors~\cite{Blatt:1952,Manley:1984jz},
\begin{align}\label{BW_fctr}
&B_0(r)=1\,,\\
&B_1(r)=r/\sqrt{1+r^2}\,,\nonumber\\
&B_2(r)=r^2/\sqrt{9+3r^2+r^4}\,,\nonumber\\
&B_3(r)=r^3/\sqrt{225+45r^2+6r^4+r^6}\,,\nonumber\\
&B_4(r)=r^4/\sqrt{11025+1575r^2+135r^4+10r^6+r^8}\, .\nonumber
\end{align}
The new free parameters $\lambda$ need to be determined from a fit to the data. For simplicity and to keep the number of parameters low, the $\lambda$s are chosen as channel-independent.
Note that one could further try to use baryon chiral perturbation theory to constrain the amplitudes at
low momenta and energies, however, the framework for doing that has not been worked out in all
necessary details~\cite{Steininger:1996xw}.

In summary, for every partial wave, the multipoles $E^\mu$, $M^\mu$ and $L^\mu$ are fully determined up to: (1) $(1+N)$ channel-dependent fit parameters $\beta^0_\mu,...,\beta^N_\mu$ for the non-pole part; (2) $(1+N)$ channel-independent parameters $\delta^0_i,...,\delta^N_i$ for each of the $i_\text{max}$ resonances; (3) one channel-independent threshold behavior regulating parameter $\lambda$; (4) channel-(in)dependent normalization factors $\zeta^{NP}_\mu(\zeta_i)$. Finally, any observable can be constructed from the described multipoles using a standard procedure involving CGLN and helicity amplitudes~\cite{Chew:1957tf}. For explicit formulas  we refer the reader to the previous publication~\cite{Mai:2021vsw}.

\section{Data and fits}
\label{sec:expdata}

In the present approach we extend the partial-wave basis to S-, P-, D- and F-waves which is necessary having extended the maximal energy range $W_{\rm max}=1.6\mapsto 1.8~{\rm GeV}$ with respect to the previous works~\cite{Mai:2021aui,Mai:2021vsw}. Including then all parameters for $\pi N,\eta N, K\Lambda$ channels, while limiting $N=2$, and fixing $\zeta^{NP}_{\mu\neq\eta N,K\Lambda}\equiv \zeta^{NP}_{\pi N}$ as well as $ \beta_{\mu\notin\{\pi N,\eta N, K\Lambda\}}^{i\in\{0,1,2\}}=0$ we obtain 533 free parameters of the model.  These parameters are fixed to the database consisting of $N_{\rm data}=110281$ data as presented in Fig.~\ref{fig:DATA}. Specifically, covered are all available electroproduction data as of 2022 in the energy region $W\in[1.13,1.8]~{\rm GeV}$ and $Q^2\in(0,8]~{\rm GeV^2}$ for the 
$$\aleph=\{\pi^0p, \pi^+n, \eta p, K^+\Lambda\}$$ 
final states. The data base covers 11 types of observables, see Fig.~\ref{fig:DATA} and Ref.~\cite{Mai:2021vsw} for explicit expressions in terms of the multipoles $\{E,M,L\}$.
In previous fits~\cite{Mai:2021aui, Mai:2021vsw} the respective solutions were used to identify many outliers due to typos in older data bases, which are cleaned up in this version and are also available through the JBW web-page~\cite{JBW-homepage}.

Fits were performed utilizing high-performance computing resources at The George Washington University~\cite{Colonial-One}. In that, the MINUIT library was used to minimize either the regular (unweighted) $\chi^2$-function
\begin{align}
&\chi^2=\sum_{i=1}^{N_{\rm all}}\left(\frac{O_i^{\rm exp}-O_i}{\Delta_i^{\rm stat}+\Delta^{\rm syst}_i}\right)^2\,,
\label{eq:chi2tot}
\end{align}
or, taking into account the very different number of data points in the $\pi N$ channels of order $\mathcal{O}(10^5)$ to those in $K\Lambda$ or $\eta N$ channels of order $\mathcal{O}(10^3)$, the weighted $\chi^2$-function
\begin{align}
&\chi^2_{\rm wt}=
\sum_{j\in\aleph} \frac{N_{\rm all}}{4N_j}
\sum_{i=1}^{N_j}\left(\frac{O_{ji}^{\rm exp}-O_{ji}}{\Delta_{ji}^{\rm stat}+\Delta^{\rm syst}_{ji}}\right)^2 \ ,
\label{eq:chi2dem}
\end{align}
where $N_j$ is the number of data for a given final state $j$.
In both of these cases, statistical and systematic uncertainties have been added linearly. We note that while this is only one possible choice, the available data base is quite heterogeneous and has consistency issues, some examples of which were discussed in Ref.~\cite{Mai:2021aui}.

Using different parameter sets determined in Ref.~\cite{Mai:2021aui} as starting values and different strategies we found two local minima for each version of the $\chi^2$-function. The results are quoted in Tab.~\ref{tab:FIT-RESULTS}.

\begin{table}[t]
\begin{tabular}{l||l|llll}
    ~~~~~~~~~~~~~&$\chi^2_{\rm dof}$~~~~&$\chi^2_{\rm pp}(\pi^0p)$&$\chi^2_{\rm pp}(\pi^+n)$&$\chi^2_{\rm pp}(\eta p)$&$\chi^2_{\rm pp}(K^+\Lambda)$\\
    \hline
    ${\bf FIT}_1$ &1.42&1.40&1.47&1.49&0.70 \\
    ${\bf FIT}_2$ &1.35&1.38&1.35&1.40&0.58 \\
    &$\chi^2_{\rm wt, dof}$~~~~&$\chi^2_{\rm pp}(\pi^0p)$&$\chi^2_{\rm pp}(\pi^+n)$&$\chi^2_{\rm pp}(\eta p)$&$\chi^2_{\rm pp}(K^+\Lambda)$\\
    \hline
    ${\bf FIT}_3$ &1.12&1.44&1.61&1.08&0.33\\
    ${\bf FIT}_4$ &1.06&1.42&1.44&1.09&0.32\\
\end{tabular}
\caption{Fit results of the present analysis with respect to standard~\eqref{eq:chi2tot} and weighted~\eqref{eq:chi2dem} ('wt')  $\chi^2$ functions. The last four columns separate out contributions for individual final-state channels ($\chi^2$ per datum) for convenience.}
\label{tab:FIT-RESULTS}
\end{table}

\begin{figure}[thb]
    \centering
    \includegraphics[width=0.95\linewidth]{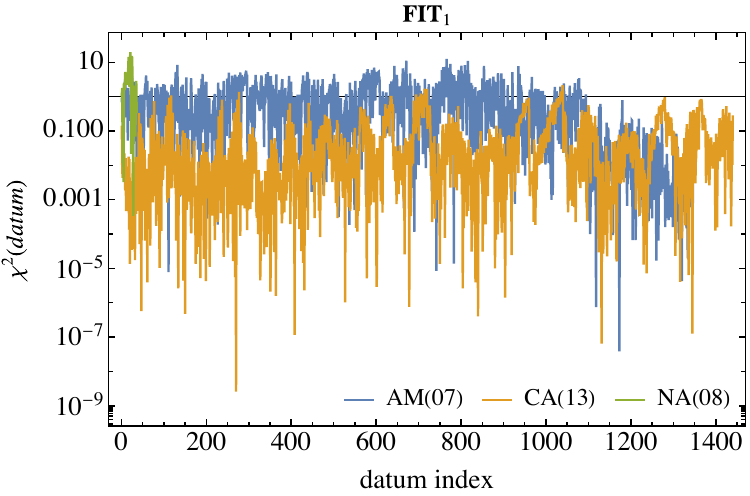}
    \caption{Distribution of partial $\chi^2$ values per datum for the $K^+\Lambda$ final states, for ${\bf FIT}_1$. Data is taken from CA(13)~\cite{Carman:2012qj}, AM(07)~\cite{CLAS:2006ogr} NA(08)~\cite{CLAS:2008agj}.}
    \label{fig:chidatum}
\end{figure}

We observe that all four solutions lead to a similar data description with the weighted solutions (${\bf FIT}_3, {\bf FIT}_4$) improving the $\eta N$ and $K\Lambda$ data description. This is indeed expected as those data have more weight in these fits.
In Fig.~\ref{fig:chidatum} we also show the $\chi^2$ for each data point color-coded for the three data references. The curves correspond to ${\bf FIT}_1$, representative for all four solutions. The most modern (2013, orange) data~\cite{Carman:2012qj} are described consistently better than the slightly older data from Ref.~\cite{CLAS:2006ogr} (blue) and Ref.~\cite{CLAS:2008agj} (green). 

\begin{figure}[thb]
    \centering
    \includegraphics[width=0.95\linewidth]{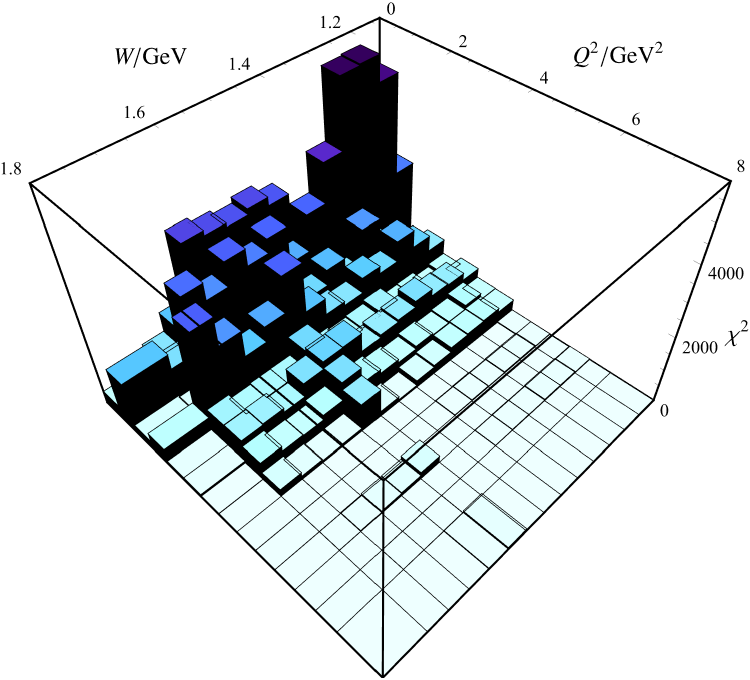}
    \includegraphics[width=0.48\linewidth,trim=0 0 1.1cm 0,clip]{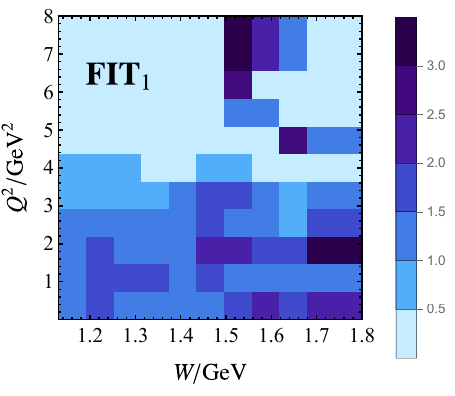}
    \includegraphics[width=0.48\linewidth,trim=0 0 1.1cm 0,clip]{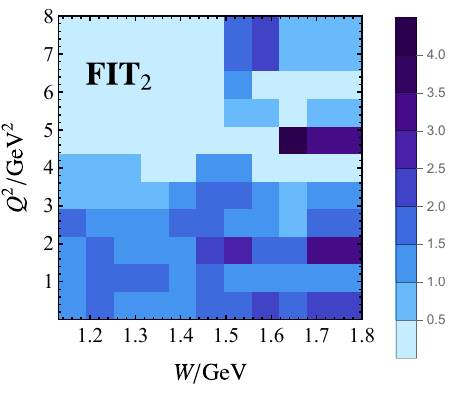}
    \includegraphics[width=0.48\linewidth,trim=0 0 1.1cm 0,clip]{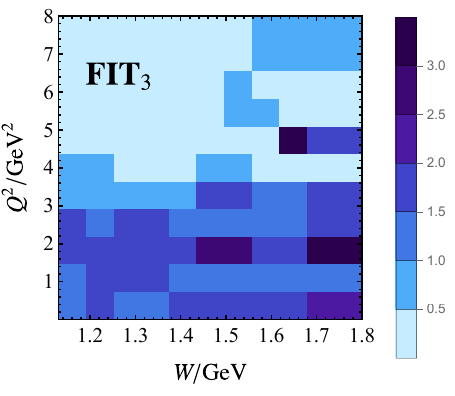}
    \includegraphics[width=0.48\linewidth,trim=0 0 1.1cm 0,clip]{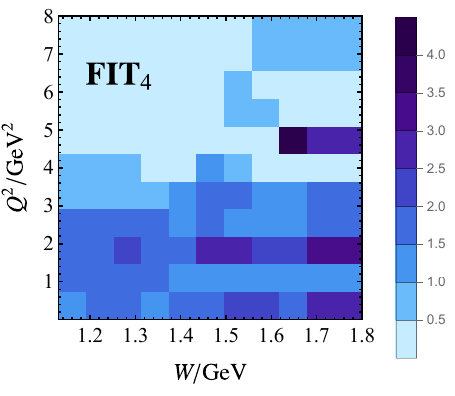}
    \caption{Distribution of $\chi^2$. Top:  
    Binned distribution of total $\chi^2(W,Q^2)$ for a typical solution, here ${\bf FIT}_1$. Bottom: Binned distribution of $\chi^2(W,Q^2)$ weighted per number of points in each $(W,Q^2)$ bin. Shades of blue correspond to the linearly scaled range of values $[0.0,4.0)$.}
    \label{fig:Chi2-3D}
\end{figure}

Furthermore,  throughout all four solutions of the present analysis, we observe that the largest contributions to the $\chi^2$ indeed come from low-$Q^2$ and low-$W$ values. This is visualized in the top panel of  Fig.~\ref{fig:Chi2-3D}. Comparing this with the data representation of Fig.~\ref{fig:DATA} the observed accumulation at low $Q^2$ and $W$ corresponds to the fact that most data are measured in that region. Still, one can also conclude that more data in the large $Q^2$ region would be very desirable. Normalizing the same binned (in $W$ and $Q^2$) $\chi^2$ distribution we obtain the bottom panels of Fig.~\ref{fig:Chi2-3D}. Discrepancies are mostly observed at higher energies, which could be a sign that G-waves become important. Also, there is a large discrepancy at $W\sim 1.6$~GeV and large $Q^2$ that comes from the difficulty to describe the $\eta N$ electroproduction data~\cite{Dalton:2008aa} in that region. In our previous analysis~\cite{Mai:2021aui} the data was not included due to the restricted $Q^2$ range, but now the large contribution to the $\chi^2$ shows that the asymptotic $Q^2$ behavior might need to be explored further in future updates of the model. In any case, ${\bf FIT}_3$ and 4 perform quite well in that region.

\section{Discussion}
\label{sec:results}

This work is the next step on the quest of uniting the description of meson-, real photon- and virtual photon-induced reactions through the dynamical coupled-channel approach. In that, several theoretical challenges have been overcome, such as including higher partial-waves (up to F-waves), extending the parameterization to independent $Q^2$-parameterization in the $K\Lambda$ channels, extending the kinematic range and, consequently, the data base. 

\begin{figure}
    \centering
    \includegraphics[width=\linewidth]{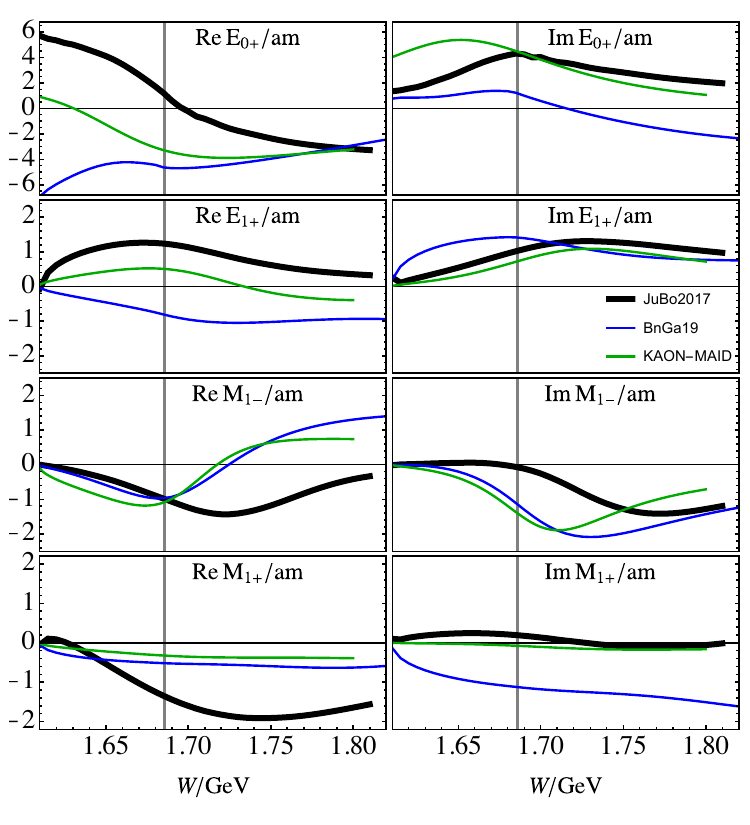}
    \caption{Model input in the $K^+\Lambda$ final state at the photo point $Q^2=0\,\GeV^2$  provided by the J\"uBo2017 solution~~\cite{Ronchen:2018ury} (black lines) compared to the Bonn-Gatchina 2019 analysis~\cite{CBELSATAPS:2019ylw} (blue) and KAON-MAID~\cite{Bennhold:1999mt,Lee:1999kd} (green, multiplied by $(-1)$ because of the different conventions). The vertical lines show the position of the $K\Sigma$ threshold while ${\rm am}$ refers to $\rm mfm$.}
    \label{fig:photo-comparison}
\end{figure}

As a first step it is useful to examine the photo-production solution which is used as input for the current analysis, i.e., J\"uBo2017~\cite{Ronchen:2018ury}. A comparison of this to other available solutions such as Bonn-Gatchina 2019~\cite{CBELSATAPS:2019ylw} or KAON-MAID~\cite{Bennhold:1999mt,Lee:1999kd} is depicted for some representative multipoles in Fig.~\ref{fig:photo-comparison}. We observe larger deviations between the models compared to the case of $\eta N$ final states (see Fig.~2 in Ref.~\cite{Mai:2021aui}). The reason is that the approaches are parametrized differently. In addition, existing data in photoproduction are not complete to uniquely pin down multipoles up to a global phase, and J\"uBo and Bonn-Gatchina fit slightly different data bases. However, except for KAON-MAID, the approaches describe the bulk of modern cross section and polarization data to a very comparable accuracy.

\begin{figure*}[thb]
    \includegraphics[width=\linewidth]{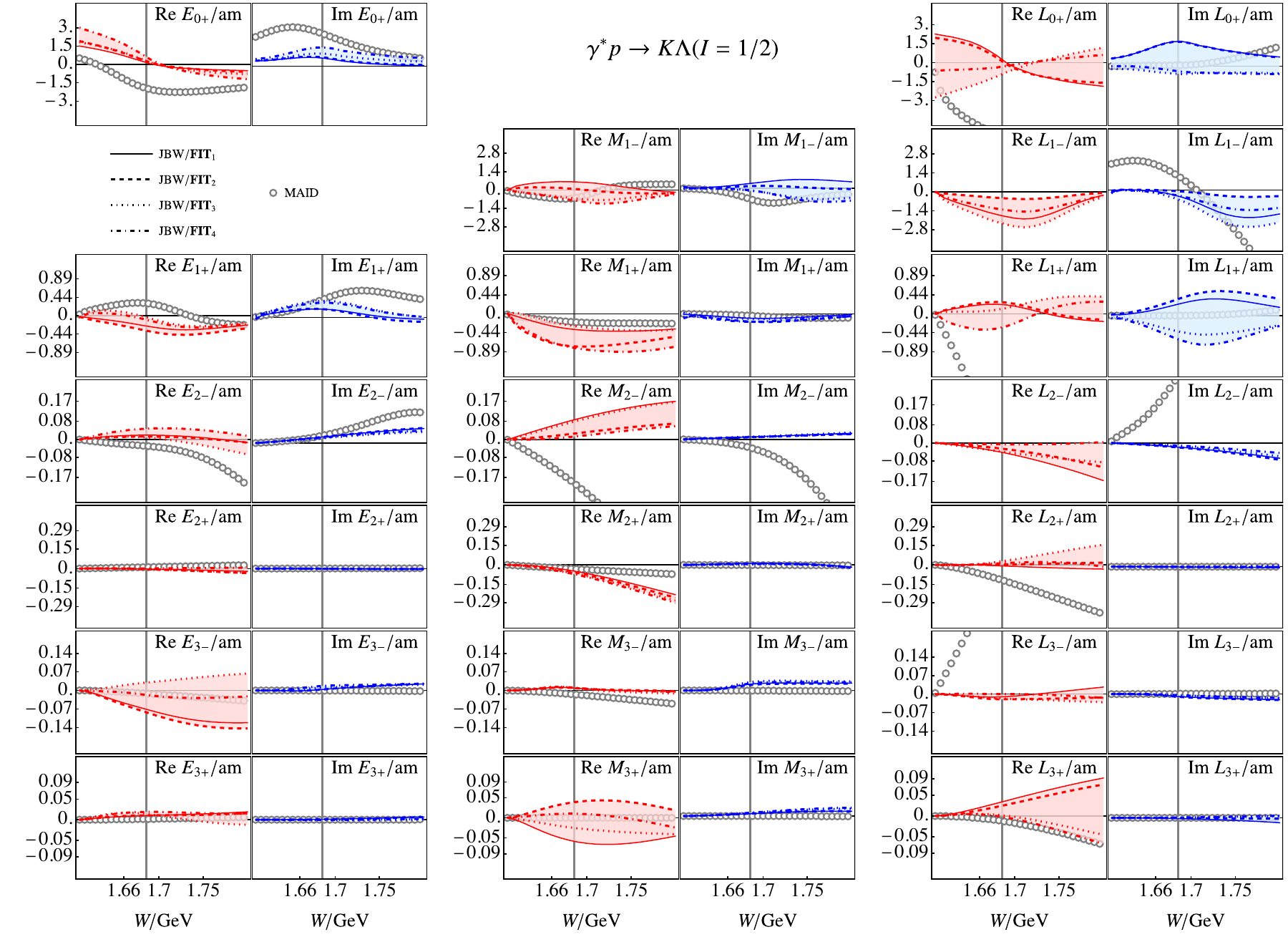}
    \caption{JBW-multipoles for the $K\Lambda\,(I=1/2)$ final state. Different solutions are denoted by ${\rm JBW}/{\bf FIT}_{1..4}$ (connected by shading to guide the eye), while circles represent the KAON-MAID~\cite{Bennhold:1999mt} solution (multiplied by $(-1)$). Results are separated w.r.t. type, angular momentum, real and imaginary parts, and are shown for a fixed value of photon virtuality $Q^2=0.5~{\rm GeV}^2$. Results for other kinematics can be obtained from the JBW web page \url{https://jbw.phys.gwu.edu}. The vertical lines shows the position of the $K\Sigma$ threshold.}
    \label{fig:ELMQ205GeV2}
\end{figure*}

\begin{figure*}[thb]
    \centering
    \includegraphics[width=\linewidth]{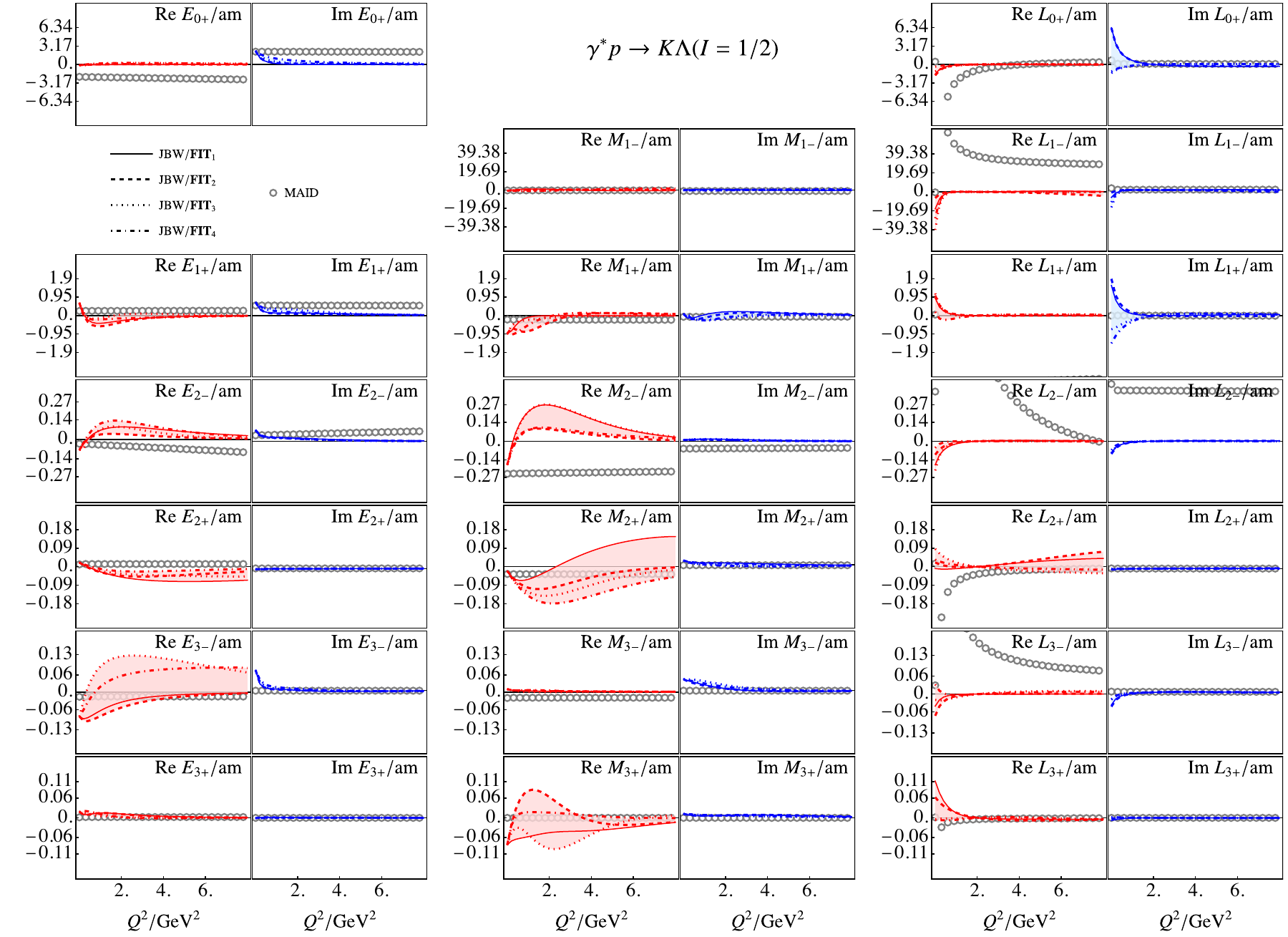}
    \caption{Multipoles for the $K\Lambda (I=1/2)$ final state  obtained through the JBW coupled-channel analysis of electroproduction data in the $\pi N, \eta N, K\Lambda$ channels. Total energy is fixed to $W=1.7~{\rm GeV}$, different solutions are denoted by ${\rm JBW}/{\bf FIT}_{1..4}$ (connected by shading to guide the eye), results of KAON-MAID~\cite{Bennhold:1999mt} are depicted by gray circles (multiplied by $(-1)$). Results at other kinematics can be obtained from the JBW web page \url{https://jbw.phys.gwu.edu}.}
    \label{fig:ELMW17GeV}
\end{figure*}

Turning now to non-vanishing virtuality, we note first that the description of the $\pi N$ electroproduction channels improved in the present study, irrespective of the utilized form of the $\chi^2$-function, according to $\chi^2_{\rm dof}\approx(1.7 \to 1.4)$ comparing to previous $\pi N$ and $\pi N/\eta N$ analyses~\cite{Mai:2021vsw,Mai:2021aui}. The obvious reason for this is the increased number of free-parameters due to the $K \Lambda$ channel, and inclusion of higher partial waves. Still, this observation is non-trivial as the number of included data has been increased as well, covering a larger kinematic range. Being more specific, we compare the estimated multipoles with those of the previous solution~\cite{Mai:2021aui}, where only S-/P-/D- waves and $\pi N/\eta N$ data in the range $W<1.6\,\GeV$ were included. We find that both $\pi N$ and $\eta N$ channels agree with the previous results while the discrepancy among the multipoles across the  four different solutions seems to have been reduced in the new result. As an explicit example, we show in the appendix (see Fig.~\ref{fig-app:piN1535} and Fig.~\ref{fig-app:etaN1535}) the $Q^2$ behaviour of the $\pi N$ and $\eta N$ multipoles projected to isospin $I=1/2$ for fixed total energy $W=1.535\,\GeV$. This is to be compared with the Fig.~7 in Ref.~\cite{Mai:2021aui}. We note, that the reduction of the differences between multipoles is somewhat indicative at this point due to the so far incomplete error-analysis, but it does make sense as much more data have now been included into the data base.
A similar behavior was found in the context of Chiral Unitary Models: In Ref.~\cite{Lu:2022hwm} the first NNLO analysis was performed that has a substantially larger number of parameters than at NLO. However, it was also the first analysis to include data from all strangeness sectors $S=-1,0,+1$ in meson-baryon dynamics simultaneously. As a result, the  uncertainties of resonance pole positions were reduced, compared to NLO analyses, despite the larger number of fit parameters.

Multipoles projected to the $K^+\Lambda$ final state are depicted in Fig.~\ref{fig:ELMQ205GeV2} for fixed $Q^2=0.5\,\GeV^2$ and those for a fixed $W=1.7\,\GeV$ in Fig.~\ref{fig:ELMW17GeV}. Results at other kinematics can be obtained from JBW web page. In these figures we also make a comparison with the results of the  KAON-MAID~\cite{Bennhold:1999mt} study. One notes directly, that agreement to this analysis can be found only in few single cases or in very restricted kinematical ranges even when taking into account possible phase convention difference between JBW and KAON-MAID. Still, it has to be noted that KAON-MAID was only fitted to a very limited subset of today's electroproduction data and should, thus, be taken with a grain of salt. In particular, it seems that the phenomenology of the KAON-MAID solutions necessarily leads to very large longitudinal multipoles. As it will be discussed below, this will have some important consequence when addressing new polarisation transfer data~\cite{CLAS:2022yzd} from the CLAS collaboration.

\begin{figure*}[thb]
    \centering
    \includegraphics[width=0.3\linewidth]{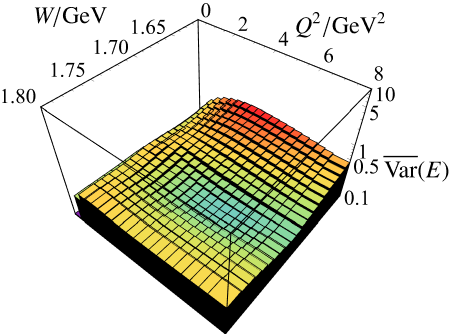}
    \includegraphics[width=0.3\linewidth]{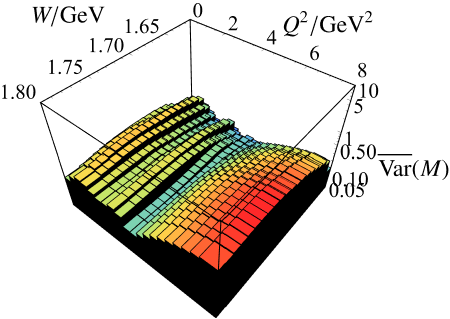}
    \includegraphics[width=0.3\linewidth]{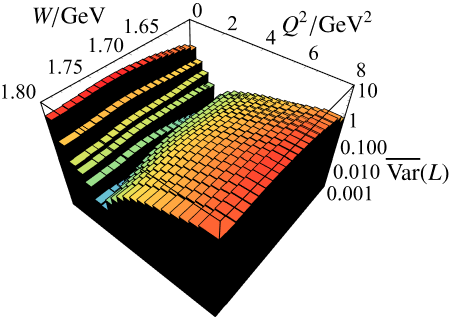}
    \includegraphics[width=0.45\linewidth]{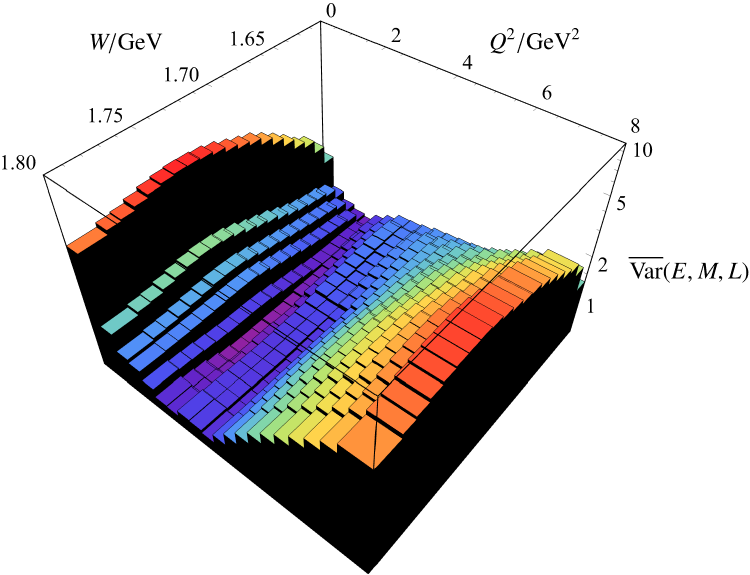}
    \caption{Systematic uncertainty of the obtained multipoles in the $K^+\Lambda$ channel binned in $Q^2$ vs. $W$ with respect to individual multipoles $\{E,M,L\}$ aggregated over total angular momentum quantum numbers. For definition of $\overline{\rm Var}(EML)$ see main text. Bottom figure shows values aggregated additionally over multipole types, i.e., $\overline{\rm Var}(EML)$.}
    \label{fig:Volatility-3D}
\end{figure*}

Comparing our obtained solutions among each other and also to the available KAON-MAID~\cite{Bennhold:1999mt} results (see, e.g., Fig.~\ref{fig:ELMQ205GeV2}) we note large theoretical uncertainty in several multipoles. In the present, largely data-driven, approach this uncertainty simply reflects the lack of data in certain kinematical regions as well as their incompleteness regarding the so-called complete experiment~\cite{Tiator:2017cde, Wunderlich:2021xhp}. 

To investigate this volatility in a more quantitative way, we define the following procedure, similar to what was proposed for the photoproduction case in Ref.~\cite{Anisovich:2016vzt}. First the kinematic range $(W/{\rm GeV},Q^2/{\rm GeV^2})\in([1.13,1.8],(0,8])$ is split up in bins. Then for each $(W,Q^2)$ bin a modified variance of all four obtained solutions $(i=1,\dots,4)$ is calculated as 
\begin{align}
    \overline{\rm Var}(W,Q^2)(X):=\sum_{\ell\pm}\frac{\rm Var\{|X_{\ell\pm,i}|\}}{\rm Mean\{|X_{\ell\pm,i}|\}+\varepsilon},
\end{align}
where $X\in\{E,M,L\}$ denotes the considered multipole type and $\varepsilon=10^{-3}\,\rm am$ is a regulator to avoid division by zero in certain cases. The result is shown for all multipoles separately in the top row of Fig.~\ref{fig:Volatility-3D}. We observe that the electric multipole is constrained quite well in nearly all kinematic regions with the highest uncertainty provided in the threshold region at moderate $Q^2$. The magnetic multipole is unrestricted only for high $Q^2$ values. The volatility of electric and magnetic multipoles is, however, dwarfed by that of the longitudinal multipole. Indeed, it shows large volatility in all kinematic regions except of a $1\lesssim Q^2/{\rm GeV^2}\lesssim3$ valley. This maybe because of the availability of the experimental data on the $K^+\Lambda$, see Fig.~\ref{fig:DATA}. Combined together $\overline{\rm Var}(E,M,L):=\overline{\rm Var}(E)+\overline{\rm Var}(M)+\overline{\rm Var}(L)$ the aggregated measure of volatility is provided in the bottom part of the Fig.~\ref{fig:Volatility-3D}. It shows clearly that it is dominated by the uncertainty in the longitudinal multipole where the most uncertain kinematic regions are those of low $Q^2$ and those of higher $Q^2$. This can be directly related to the poor data situation in this region, emphasizing again the importance of the high-virtually experimental programs such as CLAS12~\cite{Aznauryan:2012ba, CLAS:2022yzd} or the EIC~\cite{Briscoe:2015qia, Briscoe:2021cay}.

\begin{figure*}[thb]
    \centering
    \includegraphics[width=\linewidth]{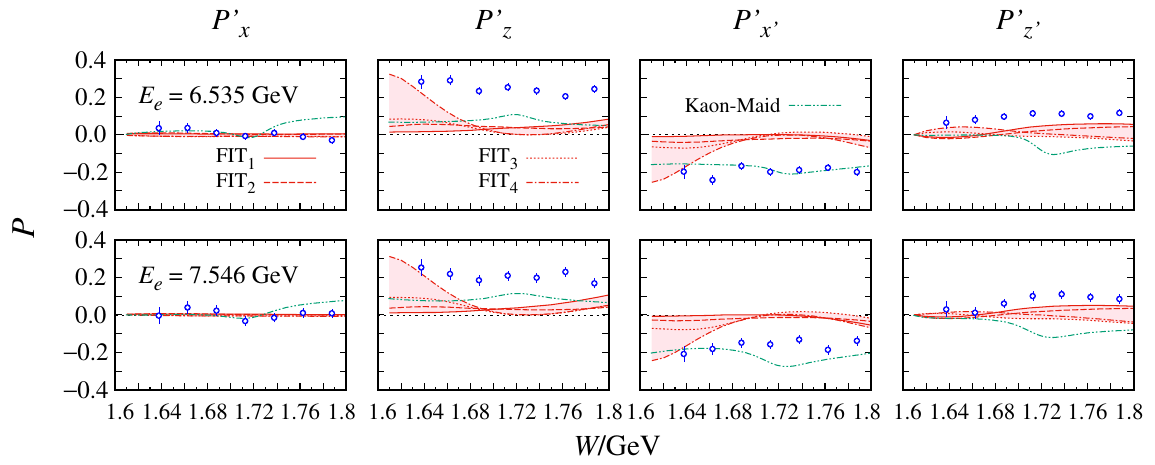}
    \caption{Transferred $\Lambda$ polarization components $P'_{x}$, $P'_{z}$, $P'_{x'}$ and $P'_{z'}$ vs. $W$ for electron beam energies of $6.535$ GeV (top panels) and $7.546$ GeV (bottom panels). Different solutions denoted by ${\rm FIT}_{1,..,4}$ (see~Tab.~\ref{tab:FIT-RESULTS}) are connected by shading to guide the eye, whereas predictions of KAON-MAID are represented by the green dash-dot-dotted lines. Experimental data are taken from Ref. \cite{CLAS:2022yzd}. }
    \label{fig:POL-transfer-CARMAN}
\end{figure*}

Having quantified that the largest uncertainties are due to the longitudinal multipoles, we proceed by calculating observables which are sensitive to these multipoles. Of particular importance is the so-called beam-recoil transferred polarization $\{P'_{a}|a=(x',z')\}$ for the Cartesian coordinate components $(x',y',z')$, such that the $e_{z'}$ is aligned with the outgoing $K^+$ and the $e_{y'}$ axis is normal to the reaction plane, see Fig.~\ref{fig:kinematics}. The corresponding quantities for the scattering plane (components $(x,y,z)$) lead to $\{P'_{a}|a=(x,z)\}$ observables, see Ref.~\cite{CLAS:2022yzd} for more details on measurement techniques and observable definitions. So far many of these data have been taken by the CLAS/CLAS12 collaboration~\cite{CLAS:2002zlc,CLAS:2009sbn,CLAS:2022yzd}, while we will refer to the most recent data~\cite{CLAS:2022yzd} from CLAS12. As such these data were taken at integrated kinematics such that $P'_{a}(W,Q^2,\cos \theta, E_e)$ is available over extended bins in one or two kinematic variables. In terms of quantities defined in the present work these \emph{integrated} observables are defined as
\begin{align}
    P'_{x'}(W,E_e)&=\frac{1}{\mathcal{N}}\int_{-1}^{+1} dc \int_{Q^2_{\rm min}}^{Q^2_{\rm max}}dQ^2 K
    \sqrt{1-\epsilon^2}R_{TT'}^{x'0}\,,\\
    P'_{z'}(W,E_e)&=\frac{1}{\mathcal{N}}\int_{-1}^{+1} dc \int_{Q^2_{\rm min}}^{Q^2_{\rm max}}dQ^2 K \sqrt{1-\epsilon^2}R_{TT'}^{z'0}\,,\\
    P'_{x}(W,E_e)&=\frac{1}{\mathcal{N}}\int_{-1}^{+1} dc \int_{Q^2_{\rm min}}^{Q^2_{\rm max}}dQ^2 K 
    \frac{\sqrt{\epsilon(1-\epsilon)}}{2}\nonumber\\
    &\hspace{3em}\left(R_{LT'}^{x'0}c-R_{LT'}^{y'0}+R_{LT'}^{z'0}s\right)\,,\\
    P'_{z}(W,E_e)&=\frac{1}{\mathcal{N}}\int_{-1}^{+1} dc \int_{Q^2_{\rm min}}^{Q^2_{\rm max}}dQ^2 K 
    \sqrt{1-\epsilon^2}\nonumber\\
    &\hspace{3em}\left(-R_{TT'}^{x'0}s+R_{TT'}^{z'0}c\right)\,,\\
    \mathcal{N}&=\int_{-1}^{+1} dc \int_{Q^2_{\rm min}}^{Q^2_{\rm max}}dQ^2 K
    \left(R_T^{00}+\epsilon R_L^{00}
    \right)\nonumber
\end{align}
where $c:=\cos\theta$, $s:=\sin\theta$, $K:=k_i'/q(Q^2=0)$, while all response functions $(R_{..}^{..})$ are functions of $(Q^2,W,c)$. Explicit form of these in terms of the multipoles can be found in Ref.~\cite{Mai:2021aui}. Note that the beam energy $E_e$ enters the right-hand side of the equations through $\epsilon(W,Q^2,E_e)$ defined in Eq.~\eqref{eq:epsilon}. The two values of $E_e$ from the Ref.~\cite{CLAS:2022yzd} measurements will be considered in the following, namely  $E_e=6.535\,\GeV$ and $E_e=7.546\,\GeV$, for which the integration limits are provided as $[Q^2_{\rm min},Q^2_{\rm max}]=[0.3,3.5]\,\GeV^2$ and $[Q^2_{\rm min},Q^2_{\rm max}]=[0.4,4.5]\,\GeV^2$, respectively. For these two cases and using all of our four solutions we postdict the results of the integrated quantities $P'_{...}(W,E_e)$, comparing them with the experimental results in Fig.~\ref{fig:POL-transfer-CARMAN}. We observe that some agreement with the data can be seen in $P'_{x}$ and $P'_{z'}$, irrespectively of the beam energy. The JBW postdictions of $P'_{x'}$ and $P'_{z}$ are, however, much smaller in the considered kinematic domain compared to the data, and they are also different from the KAON-MAID~\cite{Bennhold:1999mt} postdictions.

\begin{figure}[thb]
    \centering
    \includegraphics[width=\linewidth]{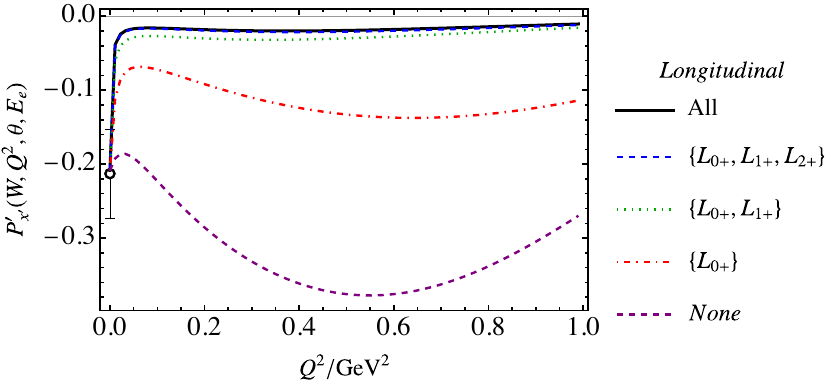}
    \caption{Polarization transfer as a function of $Q^2$ for fixed $W=2.169~{\rm GeV}$, $E_e=6.535~{\rm GeV}$ and $\theta=1.719\,{\rm rad}$. Different lines represent a ${\bf FIT}_4$ with a longitudinal multipoles included as specified in the legend. The corresponding experimental data point (black dot) is obtained from Ref.~\cite{CLAS:2006pde} by identifying $P'_{x'}=-\sqrt{1-\epsilon^2}C_{x'}$ at $Q^2=0$.}
    \label{fig:Ppxp-Q2-study}
\end{figure}

Following up on the latter discrepancy, we note that the largest differences in KAON-MAID vs. JBW results are indeed apparent for the longitudinal multipoles, see Figs.~\ref{fig:ELMQ205GeV2} and \ref{fig:ELMW17GeV}. In some cases we see an order of magnitude difference in these multipoles. The next question is now, can one identify which of the longitudinal multipoles are responsible for the stark suppression of, e.g., JBW postdicted $P'_{x'}$ values in Fig.~\ref{fig:POL-transfer-CARMAN}. To quantify this, we took one typical JBW solution $({\bf FIT}_4)$ and \emph{a-posteriori} turned off various multipoles, recalculating each time $P'_{x'}$. This is demonstrated in Fig.~\ref{fig:Ppxp-Q2-study} for $P'_{x'} (Q^2,W=2.169~\GeV,\theta=1.719~\text{rad},E_e=6.535~\GeV)$, where all except the virtuality variables are fixed to reproduce a point where experimental photoproduction data exist. Specifically, $C_{x'}$ was measured in Ref.~\cite{CLAS:2006pde} which is also included into the J\"uBo database. Identifying $P'_{x'}=-\sqrt{1-\epsilon^2}C_{x'}$ at $Q^2=0$ we indeed recover the experimental result at the photon-point. However, the predicted $P'_{x'}$ quickly goes to much smaller values when increasing $Q^2$ which we also observe for the integrated $P'_{x'}$ in Fig.~\ref{fig:POL-transfer-CARMAN}. 
We found that turning off the electric or  magnetic multipoles has little effect on this $Q^2$ behaviour. The longitudinal multipoles -- foremost the $L_{0+}$ -- can change the $Q^2$ behavior entirely. We expect, therefore, that including  polarization transfer data in a future work will have most significant impact on these multipoles.

\begin{figure}
    \centering
    \includegraphics[width=\linewidth]{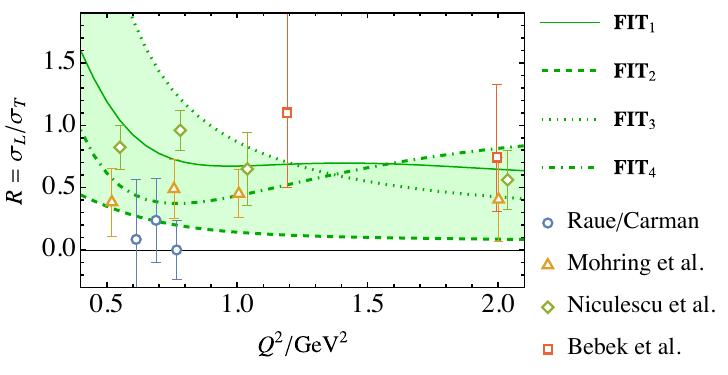}
    \caption{Ratio of $K\Lambda$ longitudinal to transverse structure functions as a function of $Q^2$. Predictions obtained from this work are provided by dark green lines  ($\theta=0, W=1.84\,\GeV$), connected by the shading to guide the eye. Experimental determinations~\cite{Raue:2004us,E93018:2002cpu,Niculescu:1998zj,Bebek:1977bv} are  depicted by empty symbols.}
    \label{fig:sigmaL/sigmaT}
\end{figure}

Finally, we compare our postdiction on the ratio of longitudinal to transverse structure functions to the experimental determinations. By construction, this ratio is strongly dependent on the longitudinal components of the transition amplitude. Additionally, there are quite a few experimental results~\cite{Raue:2004us,E93018:2002cpu,Niculescu:1998zj,Bebek:1977bv}, which have similar (but not equal) kinematics. Still, following Ref.~\cite{Raue:2004us} we compile our predictions together with experimental results in Fig.~\ref{fig:sigmaL/sigmaT}. Interestingly, our predictions seem to be well in agreement with the trend provided by the experimental determinations~\cite{Raue:2004us,E93018:2002cpu,Niculescu:1998zj,Bebek:1977bv}. One hast to note, however, that the data are less precise for this ratio than for the polarization transfer shown in Fig.~\ref{fig:POL-transfer-CARMAN}. 

\section{Summary and outlook}
\label{sec:summary}

In this work a further step has been taken towards providing a unified phenomenology of single-meson pion-induced, photo- and electroproduction data. In that, we have extended our formalism by including
up to F-waves; included a new $Q^2$ parametrization of the $K\Lambda$ channel; extended the data base and range of applicability of our formalism to $W<1.8\,\GeV$ and $Q^2<8\,\GeV^2$. Using the multipoles of the J\"uBo2017 approach at the photon-point $(Q^2=0\,\GeV^2)$ as constraint to our formalism we fit the $Q^2$ parameterization to the available experimental data $(N_{\rm data}\approx110\,000)$. 

We find that all three-channels $\{\pi N/\eta N/K\Lambda\}$ are described well. The largest source of uncertainties in the extracted multipoles comes from different local $\chi^2$ minima that we can explore due to an extensive search of the parameter space and different fitting strategies. Also, weighing the data differently in the $\chi^2$ produces large changes in extracted multipoles. This reflects the presence of substantial kinematic gaps in the data to pin down the solution, calling for more measurements. We identify the kinematic regions in which it would be most valuable to have more data. Discrepancies among extracted multipoles also reflect the presence of ambiguities due to the absence of complete-experiment coverage by different observables. The uncertainties from these sources are much larger than the effects from statistical and systematic errors of the data themselves. 

We obtain relatively good $\chi^2$ values for this type of analysis ($\chi^2_\text{dof}\approx 1.4$), but, even then, the values are not even close to being acceptable in a statistical sense. This indicates that our 500-parameter fit is not flexible enough and/or there are underestimated inconsistencies in the data base, which is a notorious problem in baryon spectroscopy.

We also observe that the discrepancies become smaller among the extracted $\{\pi N/\eta N\}$ multipoles, corresponding to different local $\chi^2$ minima, when compared to previous studies in which only the $\{\pi N/\eta N\}$ electroproduction channels were fitted. This occurs despite the larger parameter space of the current solution that includes also $K\Lambda$ electroproduction. This behavior is likely a sign that a global analysis indeed provides valuable constraints through coupled-channel effects. 

We observe 
and quantify that the largest uncertainty of our solutions is given by that of the longitudinal multipoles. This means that the available data base consisting in the $K\Lambda$ channel of cross sections only is not restrictive enough for single multipoles. Future progress can be achieved by including the recent polarization transfer data measured by the CLAS collaboration. Indeed, we checked that there is some tension between our solutions and available integrated polarization transfer observables. We showcased that low angular momentum longitudinal multipoles are the most crucial contributors to this discrepancy. 

We plan to explore the $Q^2$-dependence of the $K\Sigma$ channels addressing available experimental data. This will also allow us to consistently include the CLAS polarization-transfer data in the $K\Lambda$ and $K\Sigma$ channels. It is notable that, so far, many of these data are available only in relatively large bins. Thus, an inclusion of such data into the data base would require integrations over some kinematic variables. Measurements with smaller binning are currently on the way by the CLAS collaboration~\cite{priv_carman}.
Finally, the work on extracting helicity couplings of the resonances and the implementation of model selection techniques to produce a more efficient parametrization is ongoing.

~\\
{\bf Acknowledgements} 
The authors thank Daniel Carman, Viktor Mokeev, and Igor Strakovsky for making data available as well as for inspiring discussions, and DC and VM for a careful reading of the manuscript. The work of MM, UGM and DR was supported in part by the Deutsche Forschungsgemeinschaft (DFG, 
German Research Foundation) through the funds provided to the Sino-German Collaborative
Research Center TRR110 “Symmetries and the Emergence of Structure in
QCD” (DFG Project ID 196253076 - TRR 110). The work of UGM was further supported by
the Chinese Academy of Sciences (CAS) President's
International Fellowship Initiative (PIFI) (Grant No. 2018DM0034) and by Volkswagen Stiftung (Grant No. 93562).
The work of TM was supported by the PUTI Q2 Grant from University of Indonesia under contract No. NKB-663/UN2.RST/HKP.05.00/2022. 
The work of MD and RW was supported in part by the U.S. Department of Energy
grant DE-SC0016582; MD's work was also supported in part by DOE Office of Science, Office of Nuclear Physics under contract DE-AC05-
06OR23177. The authors gratefully acknowledge computing time on the supercomputer JURECA~\cite{jureca} at Forschungszentrum Jülich under grant no. ``baryonspectro" that was used to produce the input at $Q^2=0$.



\bibliography{BIB,NON-INSPIRE}

\clearpage
\begin{onecolumngrid}
\appendix

\section{Multipoles at fixed $W$}
\label{app-sec:Multipoles-at-fixed-W}

\begin{figure}[thb]
    \centering
    \includegraphics[width=\linewidth]{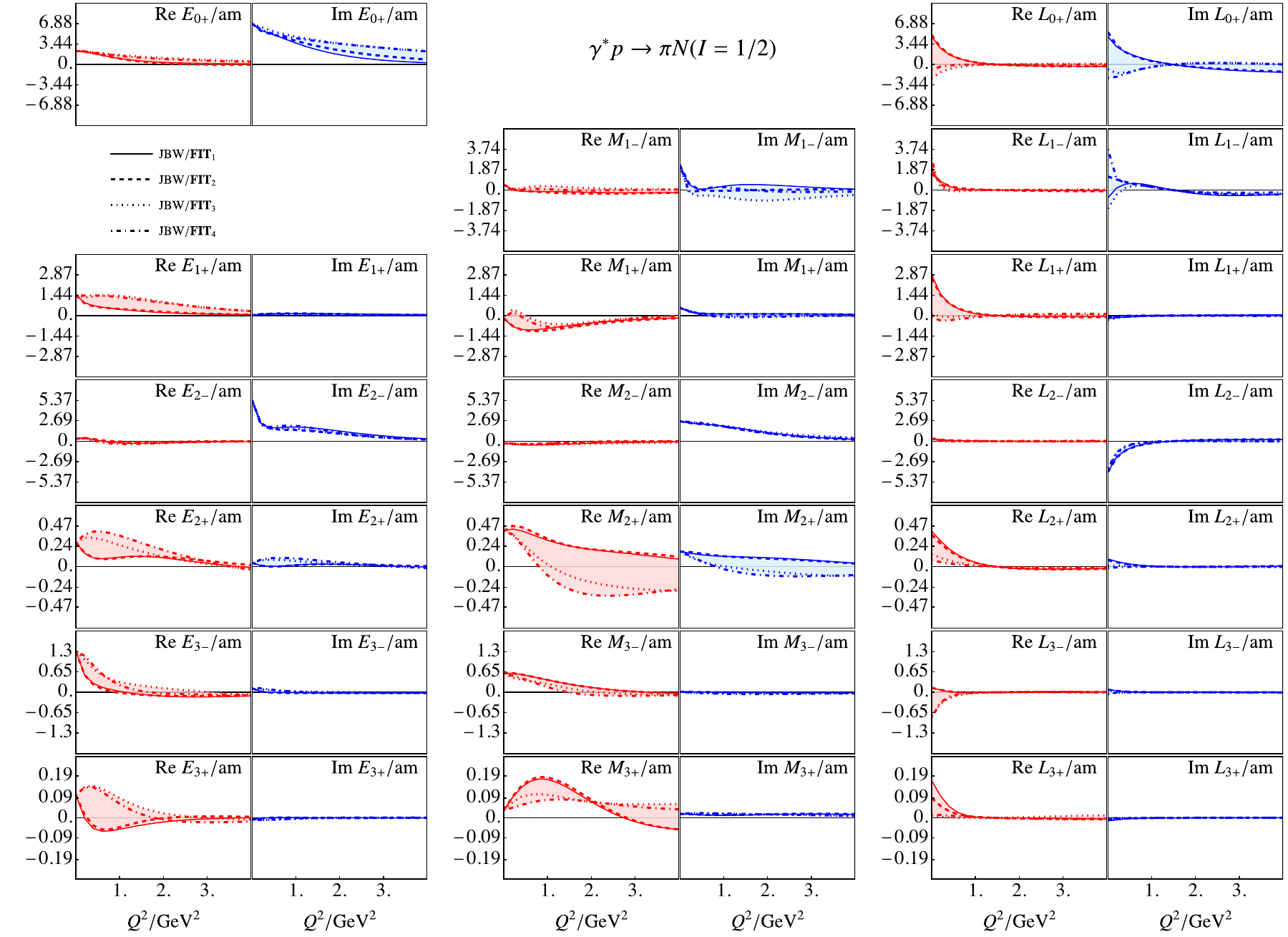}
    \caption{Multipoles for the $\pi N(I=1/2)$ final state  obtained through JBW coupled-channel solutions (connected by the shading to guide the eye)including experimental data in $\pi N, \eta N, K\Lambda$ channels. Total energy is fixed to $W=1.535~{\rm GeV}$. Results at other kinematics can be obtained from the JBW web page \url{https://jbw.phys.gwu.edu}.}
    \label{fig-app:piN1535}
\end{figure}

\begin{figure}[thb]
    \centering
    \includegraphics[width=\linewidth]{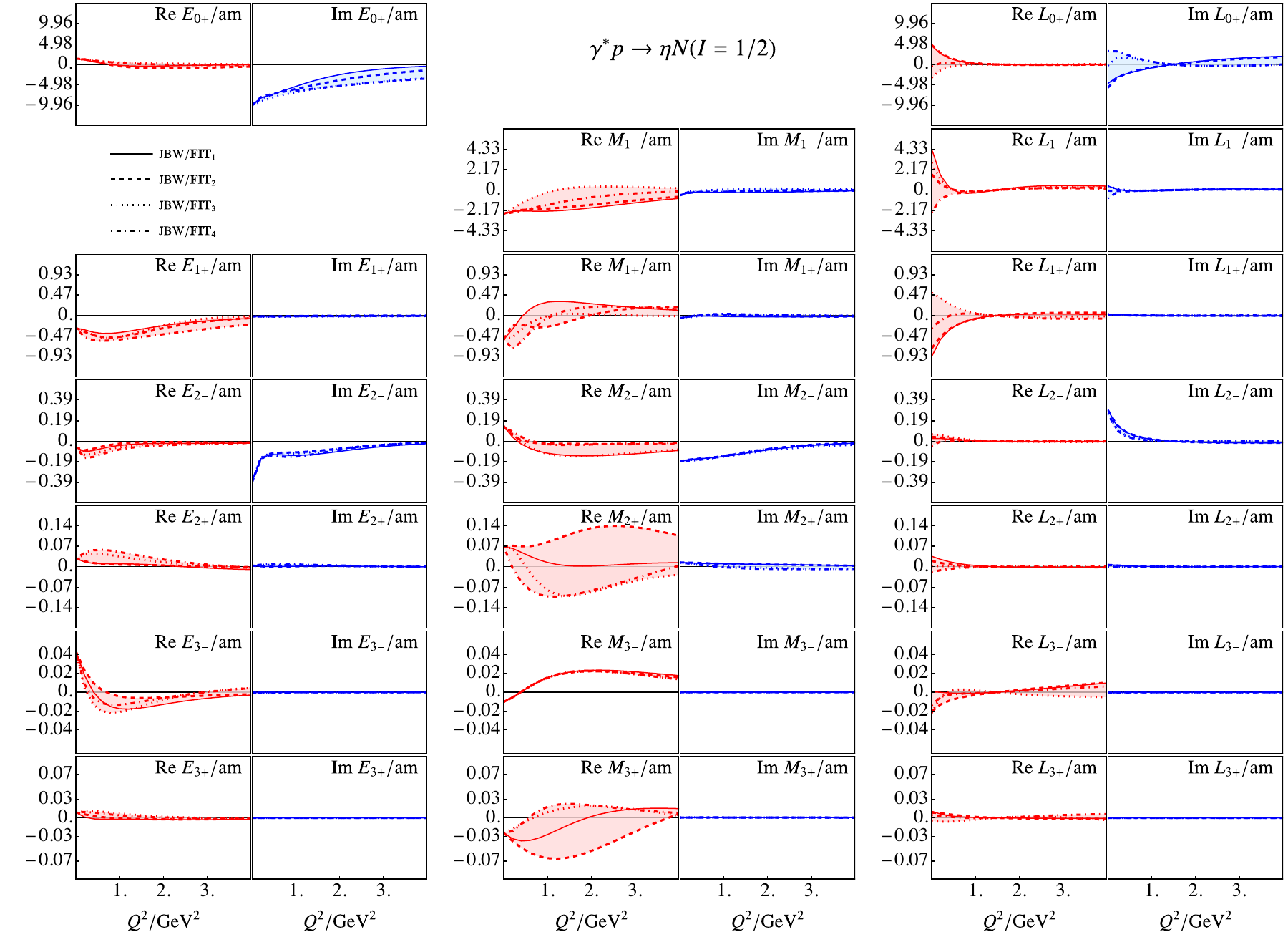}
    \caption{Multipoles for the $\eta N (I=1/2)$ final state  obtained through JBW coupled-channel solutions (connected by the shading to guide the eye) including experimental data in $\pi N, \eta N, K\Lambda$ channels. Total energy is fixed to $W=1.535~{\rm GeV}$. Results at other kinematics can be obtained from the JBW web page \url{https://jbw.phys.gwu.edu}.}
    \label{fig-app:etaN1535}
\end{figure}

\end{onecolumngrid}
\end{document}